\documentclass[twocolumn,showpacs,preprintnumbers,amsmath,amssymb,superscriptaddress]{revtex4}

\usepackage{color}
\usepackage{longtable}
\usepackage{graphicx}
\usepackage{epstopdf}
\usepackage{dcolumn}
\usepackage{bm}

\begin{document}

\title{Neutron Inelastic Scattering Processes as Background for Double-Beta Decay Experiments }
\newcommand{\lanl}{Los Alamos National Laboratory, Los Alamos, NM 87545}
\newcommand{\usd}{The Department of Earth Science and Physics, University of South Dakota, Vermillion, South Dakota 57069}
\newcommand{\uw}{Center for Experimental Nuclear Physics and Astrophysics, 
and Department of Physics, University of Washington, Seattle, WA 98195}
\newcommand{\llnl}{Lawrence Livermore National Laboratory, Livermore CA 94550}

\affiliation{	\lanl	}
\affiliation{	\usd	}
\affiliation{	\uw	}
\author{     D.-M.~Mei          }\altaffiliation[Permanent Address: ]{\usd}	\affiliation{	\lanl 	} \affiliation{	\usd 	}
\author{	S.R.~Elliott	}\affiliation{	\lanl	}		
\author{	A.~Hime		}\affiliation{	\lanl	}		
\author{	V.~Gehman	}\affiliation{	\lanl	}\affiliation{	\uw	}	
\author{    K.~Kazkaz          }\altaffiliation[Permanent Address: ]{\llnl}	\affiliation{	\uw	} 	

\date{\today}

\begin{abstract}
We investigate several Pb$(n,n'\gamma$) and Ge$(n,n'\gamma$) reactions. We measure
$\gamma$-ray production   from
Pb$(n,n'\gamma$) reactions that can be a significant background for double-beta decay
experiments which use lead as a massive inner shield. Particularly worrisome
for Ge-based double-beta decay experiments are the 2041-keV and 3062-keV
$\gamma$ rays produced via Pb$(n,n'\gamma$). The former is very close to
the $^{76}$Ge double-beta decay endpoint energy and the latter has a double
escape peak energy near the endpoint. We discuss the implications of these $\gamma$ rays
on past and future double-beta decay experiments and estimate the cross section to excite the
level that produces the 3062-keV $\gamma$ ray.
Excitation $\gamma$-ray lines from Ge$(n,n'\gamma$) reactions
are also observed. We consider the contribution of such backgrounds 
and their impact on the sensitivity of next-generation searches for 
neutrinoless double-beta decay using enriched germanium detectors.
  
\end{abstract}

\pacs{23.40.-s, 25.40.Fq, 29.40.Wk}
\maketitle

\section{Introduction}
\label{sec:Intro}
Neutrinoless double-beta decay plays a key role in understanding the neutrino's absolute mass scale and particle-antiparticle nature~\cite{Ell02, Ell04, Avi04, Bar04}. If this nuclear decay process exists, one would observe a mono-energetic line originating from a material containing an isotope subject to this decay mode. One such isotope that may undergo this decay is $^{76}$Ge.
Germanium-diode detectors fabricated from material enriched in $^{76}$Ge have established the best half-life
limits and the most restrictive constraints on the effective Majorana mass for the neutrino~\cite{igex,HM}. One analysis~\cite{kkdc} of the data in Ref.~\cite{HM} claims evidence for the decay with a half-life of $1.2 \times 10^{25}$ y.
Planned Ge-based double beta decay experiments~\cite{MJ,Gerda} will test this claim. Eventually, these future experiments target a sensitivity of $>10^{27}$ y or $\sim 1$ event/ton-year to explore mass values near that indicated by the atmospheric neutrino oscillation results. 

The key to these experiments lies in the ability to reduce intrinsic radioactive
background to unprecedented levels and to adequately shield the detectors from external
 sources of radioactivity. Previous experiments' limiting backgrounds have been trace levels of natural decay chain isotopes within the detector and shielding components. The $\gamma$-ray emissions from these isotopes can deposit energy in the Ge detectors producing a continuum, which may overwhelm the potential neutrinoless double-beta-decay signal peak at 2039 keV. Great progress has been made identifying the location and origin of this contamination, and future efforts will substantially reduce this contribution to the background. The background level goal of 1 event/ton-year, however, is an ambitious factor of $\approx$ 400 improvement over the currently best achieved background level~\cite{HM}. If the efforts to reduce the natural decay chain isotopes are successful, previously unimportant components of the background must be understood and eliminated. The potential for neutron reactions to be one of these background components is the focus of this paper. The work of Mei and Hime\cite{meihime} recognized that $(n,n'\gamma)$ reactions will become important for ton-scale double-beta decay experiments.
 Specifically, we have studied neutron reactions in Pb and Ge, materials that play important roles in the Majorana~\cite{MJ} design. But since lead is used by numerous low-background experiments, the results will have wider utility.

This paper presents measurements and simulations of Pb$(n,n'\gamma$) and Ge$(n,n'\gamma$)  reactions and estimates the resulting background for Ge-detector based, double-beta decay experiments for a given neutron flux. 
With these results, we then use the neutron flux, energy spectrum, angular distribution, multiplicity and lateral distributions determined in~\cite{meihime} to estimate the background in Ge detectors situated in underground  laboratories. In Section \ref{sec:meas} we describe the experiments, the data, and the simulations. In Sections \ref{sec:NeutronFlux} and \ref{sec:anal} we describe the analysis of these data. Section~\ref{sec:anal}  also discusses the important Pb$(n,n'\gamma$) production of $\gamma$ rays at 2041 and 3062 keV. The former is dangerously near the 2039-keV Q-value for zero-neutrino double-beta decay in $^{76}$Ge and the latter can produce a double-escape peak line at 2040 keV. These dangerous processes for Ge-based double-beta decay experiments are discussed for the first time in this work. Section \ref{sec:disc} determines an overall background model for our detector and the implications of this model for future experimental designs.
It also considers the relevant merits of Cu versus Pb  as shielding materials, and the use of  depth to mitigate these backgrounds is discussed. We also consider the possibility that the double-escape peak of the 3062-keV $\gamma$ ray could contribute to the signal claimed in Ref.~\cite{kkdc}. Finally, we summarize our conclusions in Section \ref{sec:concl}.

\section{The measurements }
\label{sec:meas}

We collected five data sets to explore the implications of $(n,n'\gamma$) for double-beta decay experiments. All
 measurements were done in our basement laboratory at Los Alamos National Laboratory. The laboratory
  building  is at an atmospheric depth of 792 g/cm$^{2}$ and provides about 1 mwe concrete (77 g/cm$^2$)
   overburden against cosmic ray muons. 
  
  Three data sets were taken with a CLOVER detector~\cite{clod}. This detector is a set of 4 n-type, segmented germanium detectors. The four
crystals have a total natural germanium mass of 3 kg and each crystal is segmented in half.  The CLOVER detector and its operation in our laboratory
were described in Ref.~\cite{ELL05}. The remaining two measurements were done with a PopTop detector~\cite{pop} set up in coincidence with
a NaI detector. The PopTop is a 71.8-mm long by 64-mm diameter p-type Ge detector. Taking into account the central bore,
the detector is 215 cm$^{3}$ or 1.14 kg. The NaI crystal is 15.25-cm long by 15.25-cm diameter 
and is directly connected to a photo-tube. All data were read out using a pair of X Ray
Instrumentation Associates (XIA) \cite{XIAref} Digital Gamma Finder 
Four Channel (DGF4C) CAMAC modules.  The CAMAC crate is connected to the PCI bus
of a Dell Optiplex computer running Windows 2000.  The system was
controlled using the standard software supplied by XIA.  This data
acquisition software runs in the IGOR Pro environment \cite{IGORref} 
and produces binary data files that were read in and analyzed using the ROOT framework\cite{ROOTref}.

  The data sets include:
  \begin{enumerate}
  \item A background run with the CLOVER 
  \item A Th-wire source run with the CLOVER  
  \item An AmBe source run with the CLOVER using two different geometries of moderator
  \item An AmBe source run with the PopTop surrounded by lead 
  \item An AmBe source run with the PopTop surrounded by copper 
  \end{enumerate}
In this section we describe the experiments and the data collected.

\subsection{The Experimental Configurations}

The CLOVER was surrounded by 10 cm of lead shielding to reduce the signal from ambient radioactivity. Underneath 
and above
the lead was 5 cm of 30\%-loaded borated polyethylene to reduce thermal neutrons. The background run
 done in this configuration lasted 27.13 live-days. The configuration for the Th-source run was similar, but with some
lead removed to expose the detector to the source. The Th source run had a live time of 1337 seconds.

\begin{figure}[htb!!!]
\includegraphics[angle=0,width=7.5cm]{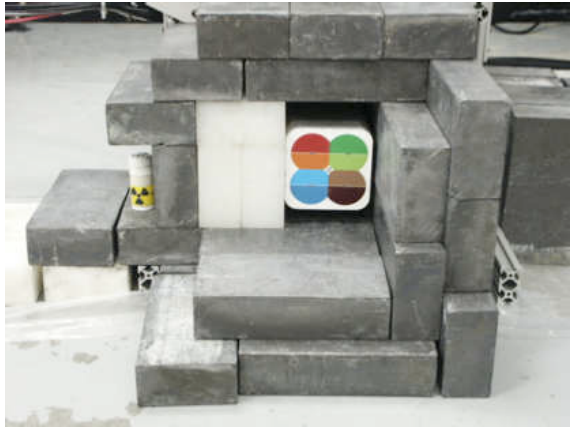}
\caption{\small {The CLOVER detector as configured for the AmBe source run. The setup at the time of this photograph used $4''$ of polyethylene.
One wall of the lead shield was removed only to clarify the relationship between the AmBe source, moderator, and the CLOVER. }}
\label{fig:setup}
\end{figure}

The setup was modified somewhat from this background-run configuration for the measurements with the AmBe source.
Fig.~\ref{fig:setup} shows the configuration for one of the AmBe measurements. For these data,  the CLOVER
 was shielded on four sides with 10 cm of lead. The AmBe source, 30 mCi of $^{241}$Am with a calibrated neutron yield of 
 $\approx$ 63,000 Hz ($\pm$0.7\%),
 was on one side of the CLOVER 
with 5 cm of lead and a layer of pure polyethylene moderator (either 10 or 15 cm thick) between the source 
and detector.  The data acquisition system is inactive during data transfer. Only the  AmBe runs had a large enough event rate for the dead time to be appreciable. A 6.13-h live-time data run (57\% live) was taken with 15 cm of
moderator, and another 3.57-h live-time data (38\% live) run was taken with 10 cm moderator (pictured).  For the analysis presented below, the data from these
two configurations were combined, and thus the AmBe-CLOVER data set contains 9.7 h of live time.
The observed energy spectrum extended from $\approx$ 10-3100 keV for these data sets. 

During the analysis of the AmBe data, we observed a weak line at 3062 keV. This energy corresponds to a $\gamma$-ray
transition in $^{207}$Pb, and we therefore hypothesized that it was generated via Pb$(n,n'\gamma$). The double-escape-peak
(DEP) energy (2040 keV) associated
with this $\gamma$ ray is very dangerous for $^{76}$Ge neutrinoless double-beta decay experiments because it falls so close to the transition
energy (2039 keV). Furthermore because the DEP is a single-site energy deposition, it cannot be distinguished from double-beta decay 
through event topology. This is in contrast to a full-energy $\gamma$-ray peak, which tends to consist of several interactions and therefore 
is a multiple-site deposition. (See ~\cite{ELL05} for a discussion of the use of event topology to reduce background in Ge detectors.)

The final two measurements were intended to study this 3062-keV line in the spectrum and demonstrate its origin.  In both cases a PopTop Ge detector faced a 15.25 cm by 15.25 cm NaI detector for coincidence data.
 By sequentially placing a Pb and then a Cu absorber between an AmBe source and a PopTop Ge detector, we tested the hypothesis that the line was due to neutron interactions in Pb. By looking for an coincident energy deposit in the NaI detector, we could be assured the Ge detector signal originated from a neutron interaction in the sample.
 An energy deposit threshold in the NaI of greater than 200 keV was required for a coincidence. The PopTop was placed 27.3 cm from the NaI detector with the source placed 20.3 cm (7 cm) from the Ge (NaI) detector. For 
the lead study, 5 cm of lead was placed directly between the Ge detector and the source. Additional lead, in the form of 5-cm-thick bricks
was positioned around the 4 sides of the Ge detector to reduce room background. For the copper study, a 0.5-cm thick Cu tube was placed around 
the PopTop and a 5-cm Cu block was placed between the PopTop and the source. For this final run, all the lead was removed. For both of these sets of 
data, the observed spectra extended from $\approx$ 125 keV to $\approx$ 9 MeV. For the
PopTop data, the Pb and Cu runs were of 19.12 h and 17.76 h live-time, respectively.

\subsection{The Data sets}

\begin{figure}[htb!!!]
\includegraphics[angle=0,width=7.5cm]{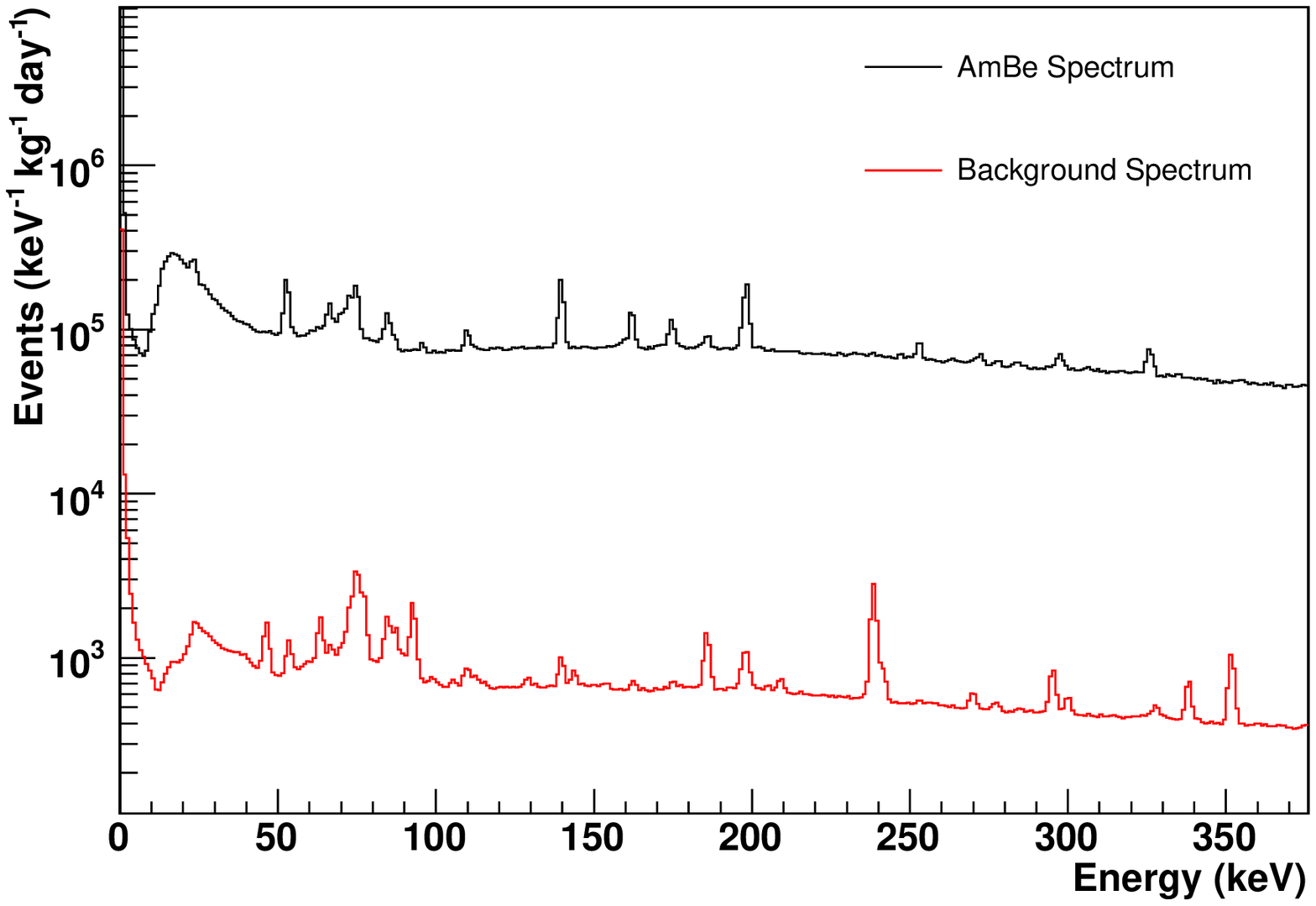}
\includegraphics[angle=0,width=7.5cm]{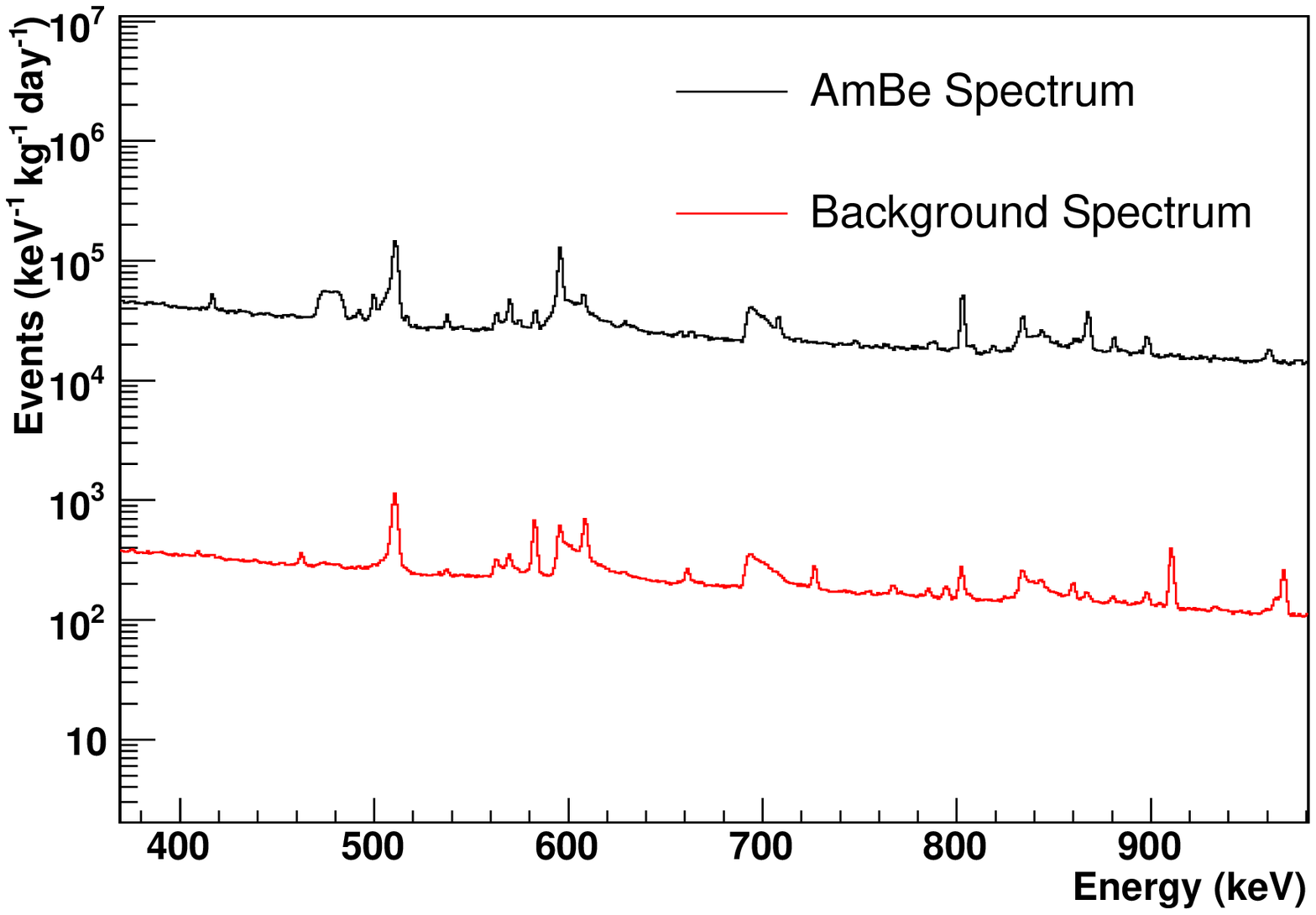}
\includegraphics[angle=0,width=7.5cm]{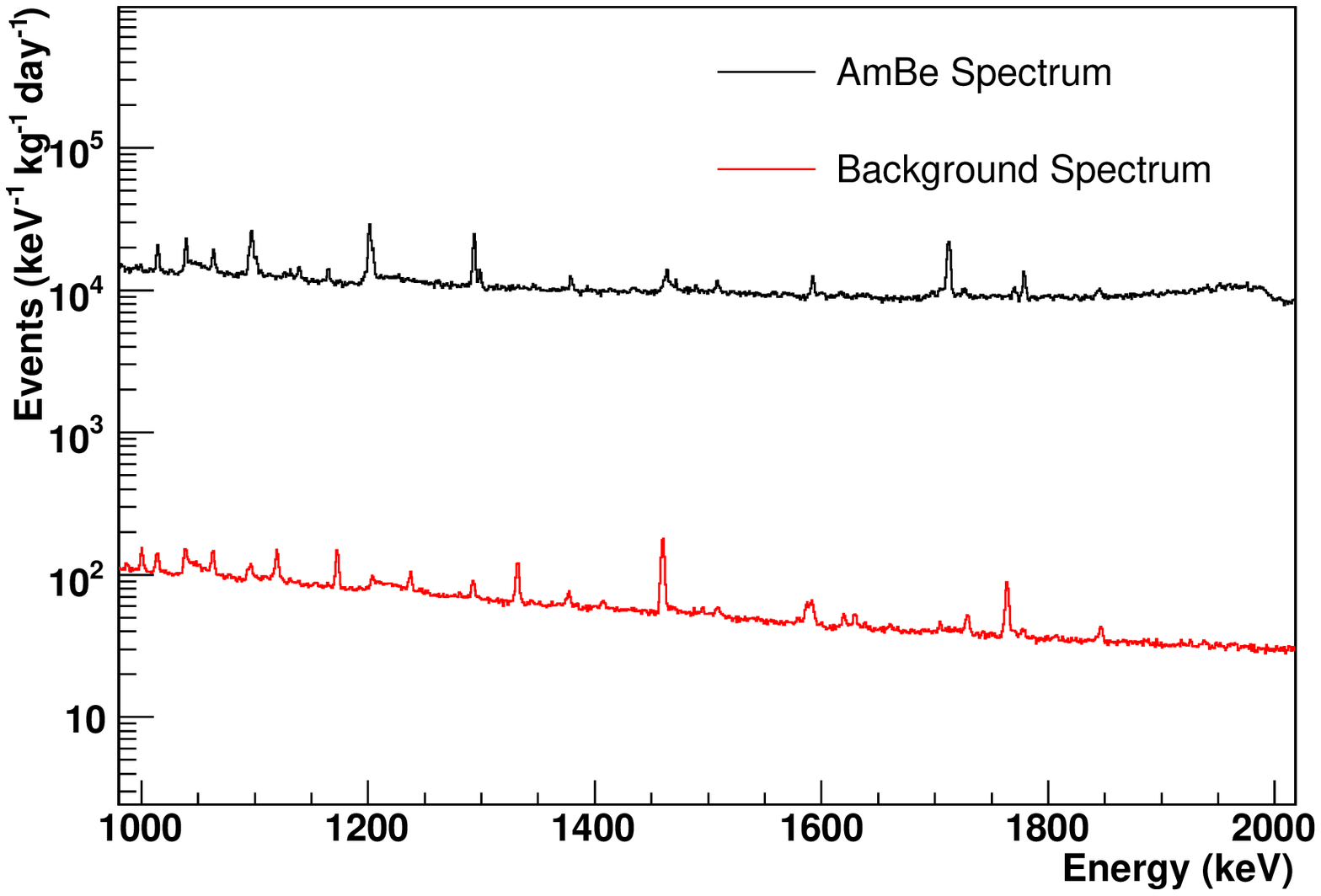}
\includegraphics[angle=0,width=7.5cm]{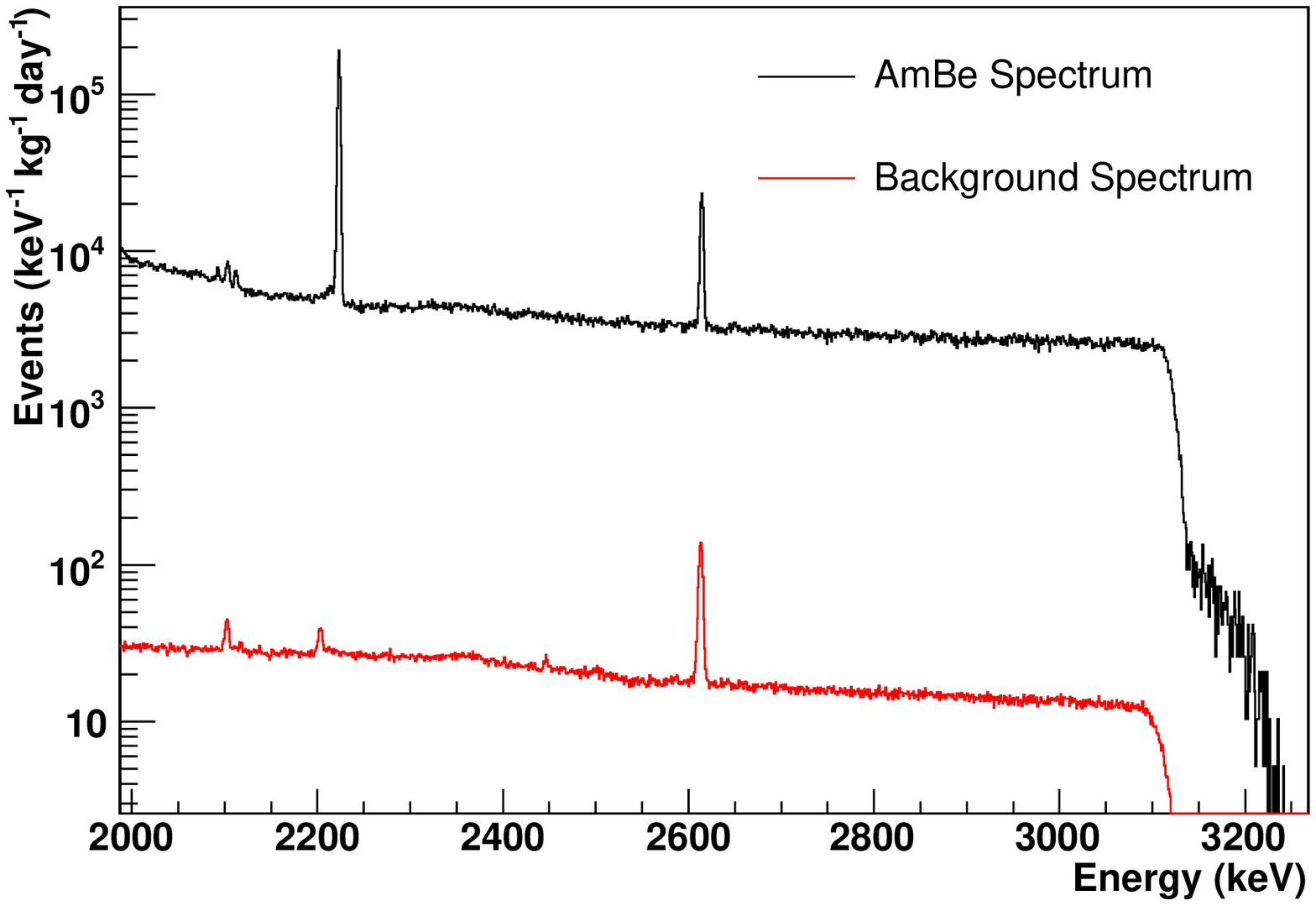}
\caption{\small {The AmBe and background spectra taken with the CLOVER. }}
\label{fig:energyspectrum1}
\end{figure}

The crystals  were individually calibrated, and the resulting spectra summed together to form a single histogram. The peaks within each of the 3 CLOVER data sets were identified and their intensities determined.  If an event had 2 crystals that responded in coincidence, the histogram would have two entries. Therefore the spectra we analyzed and simulated included all single-crystal energy deposits. By not eliminating events that registered signals in more than 1 of the CLOVER Ge detectors, we maximized the event rate. The peak strengths were estimated by fitting a Gaussian shape to peaks and a flat background to the spectrum in the region near the peak. For the nuclear recoil lines, the peak shape was assumed to be a triangle
and not Gaussian. In Table~\ref{tab:LineID} the uncertainties derive from this fit.
A summary of the peak
strengths is given in Table~\ref{tab:LineID} and the spectra themselves are shown in Fig.~\ref{fig:energyspectrum1}. The data sets were chosen to help decouple line blendings. Because the rates in all peaks and continua are much higher for the 
source-induced data than for the background, features in those spectra are due to the sources and
other contributions can be safely ignored.  For example, the 2614.5-keV
line can arise from either the decay of $^{208}$Tl or $^{208}$Pb$(n,n'\gamma)$. When exposed to a Th source, Tl decay dominates the spectrum, whereas when exposed to an AmBe source, $(n,n'\gamma$) dominates. Hence by normalizing the rate in this line to the rate in a pure neutron-induced transition (\emph{e.g.} the 596-keV $^{74}$Ge(n,n'$\gamma$)), we can 
determine the relative contribution of the two processes to the background spectrum. In fact, in the background data, both processes contribute to this line. 

Some comments on our choices for line identification are in order. For an isotope such as $^{72}$Ge where a neutron capture leads to a stable nucleus, almost all (n,$\gamma$) lines could also be interpreted as $(n,n'\gamma$) lines in the resulting nucleus; in this case $^{73}$Ge. For isotopes within the detector however, such as the 53.5-keV $^{72}$Ge(n,$\gamma$) transition, the competing 
$^{73}$Ge$(n,n'\gamma$) line would be a sum of this $\gamma$-ray energy and the recoil nucleus energy. At these low energies where the recoil is a fair fraction of the $\gamma$-ray energy, the $(n,n'\gamma$) would simply contribute to the continuum and not be observed as a line. For the high energy cases, the 
blend of a mono-energetic $\gamma$-ray line and a $(n,n'\gamma$) process might be present. 

For the calibration runs, our threshold was approximately 70 keV. Also note that we used a thorium wire as a calibration source. Since the wire is pure natural Th, we observe the Th X rays in that data. In contrast, the background run shows lines from the thorium chain as a contaminant, therefore those lines are absent.

In all spectra, there are a few lines we have not identified.


\begin{center}
\begin{longtable}{|c|c|c|c|c|}
\caption{\protect  A summary of the observed lines in the various spectra 
taken with the CLOVER detector.  Blank entries indicate that no 
significant peak feature above the continuum was found. Single and 
double escape peaks are labeled by SEP and DEP respectively.
Line assignments for which we are unsure are indicated by a
question mark. Line energies are taken from the Table of 
Isotopes~\cite{TOI}. 
}
\label{tab:LineID} \\

\hline \hline
Energy       &     Process                           &   \multicolumn{3}{c|} {Count Rates}                              \\
 (keV)         &                                               & backgrnd               &  Thorium         &  CLOVER                       \\
                   &                                              &    (per hr)                     &     (Hz)                 &  AmBe (Hz)          \\
\hline
\endfirsthead

\multicolumn{5}{c}%
{{\bfseries \tablename\ \thetable{} -- continued}} \\
\hline
Energy       &     Process                           &   \multicolumn{3}{c|} {Count Rates}                              \\
 (keV)         &                                               & backgrnd               &  Thorium         &  CLOVER                       \\
                    &                                              &    (per hr)                     &     (Hz)                 &  AmBe (Hz)          \\
\hline
\endhead

\hline 
\multicolumn{5}{c}{Continued} \\ 
\hline
\endfoot

\hline  
\hline
\endlastfoot

 23.4 & $^{70}$Ge(n,$\gamma$)            &                            &                             & 1.017(5)                                                                              \\
 46.5 & $^{210}$Pb                                  &112.75(42)        &                             &                                                                                                              \\
$\left. \begin{array}{c}53.2 \\ 53.5 \end{array} \right\} $  & $\left. \begin{array}{c}^{234}\mbox{U} \\ ^{72}\mbox{Ge(n,}\gamma\mbox{)} \end{array} \right\} $ & 61.01(31)  &   & 2.079(8) \\
 63.2 & $^{234}$Th                                  & 93.02(38)         &                             &                                                                                                              \\
 67.7 &  $^{230}$Th                                 & 24.05(19)         &                              & 0.591(4)                                                                              \\
 68.8 &$^{72}$Ge(n,$\gamma$)                &                           &                              & 0.440(4)                                                                              \\
 72.80 & Pb x-ray                                      &                           &    10.5(1)             & 0.506(4)                                                                              \\
 74.97 & Pb x-ray                                     & 663.5(1.0)         &   34.1(2)            & 1.703(7)                                                                       \\
 76.7 & Unidentified                              &                            &   29.0(2)            &                                                                                                              \\
 $\left. \begin{array}{c}84.4 \\ 84.5 \\ 84.9 \end{array} \right\} $             & $\left. \begin{array}{c}^{228}\mbox{Th} \\ \mbox{Pb x-ray} \\ \mbox{Pb x-ray} \end{array} \right\} $            
 &115.28(42)       &   10.9(1)             & 0.930(5)           \\
 $\left. \begin{array}{c}87.2 \\ 87.4 \\ 87.9 \end{array} \right\} $ & Pb x-ray              & 55.65(29)        &   16.8(1)             & 0.262(3)           \\
 89.9 &    Th   x-ray                                   &                            &   32.4(2)             &                                                                                                              \\
 92.7 & $^{234}$Th                                   & 171.38(51)     &                             & 0.060(1)                                                                              \\
 93.4 & Th   x-ray                                     &                             &  47.5(2)              &                                                                               \\
 96.0 & $^{115}$In(n,$\gamma$)    ?       &                           &                             & 0.166(2)                                                                               \\
 99.5 & $^{228}$Ac                                  & 13.71(15)        &   3.0(1)                &                                                                                                              \\
 105.3 & Unidentified                              & 8.99(12)          &                &                                                                                                              \\
 $\left. \begin{array}{c}104.8 \\ 105.6 \end{array} \right\} $   &      Th   x-ray    &           &  20.6(1)              &                                               \\
 108.7 &     Th   x-ray                                 &                           &  7.6(1)                &                                                                                                              \\
 109.9 & $^{19}$F($n,n'$$\gamma$)             & 43.00(26)       &                               & 0.506(4)                                                                              \\
  129.1 & $^{228}$Ac                                &12.91(14)       &  3.2(1)                 &                                                                                                              \\
 139.7 & $^{74}$Ge(n,$\gamma$)             & 47.20(27)      &                               & 2.339(8)                                                                              \\
 143.9 & $^{230}$Th                                & 20.03(18)      &                                &                                                                                                              \\
 154.0 & $^{228}$Ac                                & 7.69(11)         & 1.5(1)                    &                                                                                                              \\
 159.7 & $^{77m}$Ge                              &                         &                              & 0.114(2)                                                                              \\
 162.4 & $^{115}$In(n,$\gamma$)              & 10.71(13)     &                                & 1.073(6)                                                                              \\
 174.9 & $^{70}$Ge(n,$\gamma$)             & 7.45(11)        &                               &  0.763(5)                                                                              \\
  $\left. \begin{array}{c}186.1 \\ 186.2 \end{array} \right\} $  & $\left. \begin{array}{c}^{226}\mbox{Ra} \\ ^{115}\mbox{In(n,}\gamma\mbox{)} \end{array} \right\} $ & 114.60(42) &  & 0.323(3) \\
$\left. \begin{array}{c}197.1 \\ 198.4 \end{array} \right\} $  & $\left. \begin{array}{c}^{19}\mbox{F($n,n'$}\gamma\mbox{)} \\ ^{71}\mbox{Ge}\mbox{ sum} \end{array} \right\} $ & 81.04(35) &  & 2.328(8) \\
 199.2 & $^{228}$Ac                                &                          &  0.66(2)               &                                                                                                              \\
 202.6 & $^{115}$In(n,$\gamma$)       &                         &                               & 0.061(1)                                                                              \\
 209.5 & $^{228}$Ac                                 & 19.38(17)      & 10.9(1)                &                                                                                                              \\
 215.5 & $^{228}$Th                                 & 2.43(6)           & 0.92(3)                &                                                                                                             \\
 238.6 & $^{212}$Pb                                & 295.77(67)    & 139.9(1)              &0.105(2)                                                                              \\
 242.0 & $^{214}$Pb                                & 57.49(30)     &  9.2(1)                   &                                                                                                              \\
  247.1& $^{70}Ge$(n,$\gamma$)              &                        &                               &0.070(1)                                                                              \\
 253.7 & $^{74}Ge$(n,$\gamma$)              & 2.76(7)         &                               &0.410(3)                                                                              \\
270.2 & $^{228}$Ac                                & 21.39(18)      &  9.1(1)                  &                                                                               \\
 273.0 & $^{115}$In(n,$\gamma$)           &                         &                               &       0.055(1)                                                                                                       \\
 277.4 & $\left\{ \begin{array}{c}^{208}\mbox{Tl} \\ ^{208}\mbox{Pb($n,n'$}\gamma\mbox{)} \end{array} \right\} $       &12.98(14)      &  5.4(1)                   & 0.086(2)   \\
 284.6 &  Unidentified                             & 2.79(7)           &                               &                                                                                                              \\
 288.1 & $^{212}$Bi       ?                     &                        & 1.03(3)                  &                                                                                                              \\
 295.2 & $^{214}$Pb                               & 58.02(30)     &                                 &                                                                                                            \\
$\left. \begin{array}{c}297.2 \\ 298.7 \end{array} \right\} $  & $\left. \begin{array}{c}^{72}\mbox{Ge(n,}\gamma\mbox{)} \\ ^{115}\mbox{In(n,}\gamma\mbox{)} \end{array} \right\} $ 
 &                        &                               & 0.068(1)                                                                              \\
 300.1 & $^{212}$Pb                               & 18.59(17)     & 9.7(1)                    &                                                                                                              \\
 306.2 & $^{70}$Ge(n,$\gamma$)           & 1.12(4)         &                                & 0.046(1)                                                                              \\
 321.4 & $^{228}$Ac                                &                       &  0.67(2)                  &                                                                                                              \\
 326.0 & $^{70,72}$Ge($n,n'$$\gamma$)         &                       &                                 & 0.487(4) \\
 328.3 & $^{228}$Ac                               & 10.02(12)      &  8.2(1)                  &                                                                                                              \\
 332.9 & $^{228}$Ac                               &                         & 1.14(3)                 &                                                                                                              \\
 335.5 & $^{115}$In(n,$\gamma$)             &                        &                                & 0.028(1)                                                                              \\
 338.7 & $^{228}$Ac                               & 45.13(26)      & 31.5(2)                 &                                                                                                              \\
 351.9 & $^{214}$Pb                               & 95.11(38)     &                               &                                                                                                              \\
 354.1 & Unidentified                              &                        &                               & 0.043(1)                                                                              \\
 385.1 & $^{115}$In(n,$\gamma$)             &                         &                              & 0.048(1)                                                                              \\
 391.3 & $^{70}$Ge(n,$\gamma$)             &                         &                               &0.053(1)                                                                              \\
 409.8 & $^{228}$Ac                               & 4.05(8)          & 4.3(1)                    &                                                                                                              \\
 416.9 & $^{116m}$In                            & 2.21(6)           &                                & 0.359(3)                                                                              \\
 438.9 & Unidentified                             & 2.54(6)          &                               &                                                                                                               \\
 445.2 & $^{74}$Ge(n,$\gamma$)             &                         &                                 &  0.037(1)                                                                              \\
 452.3 & $^{212}$Bi?                             &                         & 0.82(2)                &                                                                                                              \\
 463.3 & $^{228}$Ac                               & 10.96(13)     &  9.2(1)                     &                                                                                                              \\
 474.0 & $^{72 }$Ge(n,$\gamma$) ?        & 2.54(6)           &                                 &                                                                                                             \\
 478.6 & $^{228}$Ac                               &                         &  0.39(2)                &                                                                                                              \\
 470-485 & $\left\{ \begin{array}{c}^{10}\mbox{B(n,}\alpha \mbox{)}^7\mbox{Li}^*  \\ ^7\mbox{Li}^*(\gamma)^7\mbox{Li} \\ \mbox{Doppler} \\ \mbox{Broadened} \end{array} \right\} $   &     &      &         signf.       \\
 492.9 & $^{73}$Ge(n,$\gamma$)        &                          &                               & 0.123(2)                                                                              \\
 499.9 & $^{70}$Ge(n,$\gamma$)         &                          &                               &0.453(4)                                                                              \\
503.9 &  $^{228}$Ac                                &                         &  0.34(2)               &                                                                                                              \\
  $\left. \begin{array}{c}509.3 \\ 510.7 \\ 510.7 \\510.9 \end{array} \right. $  & $\left. \begin{array}{c}^{228}\mbox{Ac} \\ ^{208}\mbox{Tl} \\ ^{208}\mbox{Pb($n,n'$}\gamma\mbox{)} \\ \mbox{Annih.} \gamma \end{array} \right\} $ & 171.93(51)    &  16.8(1)                & 3.409(10) \\
 516.2 & $^{35}$Cl(n,$\gamma$)  &                                    &                               & 0.160(2)                                                                              \\
 537.5 & $^{206}$Pb($n,n'$$\gamma$)       & 5.12(9)           &                               & 0.158(2)                                                                              \\
$\left. \begin{array}{c}562.9 \\ 563.0 \end{array} \right\} $  & $\left. \begin{array}{c}^{228}\mbox{Ac} \\ ^{76}\mbox{Ge($n,n'$}\gamma\mbox{)} \end{array} \right\} $ 
  & 12.83(14)       &  1.52(3) & 0.244(3)  \\
569.7 & $^{207}$Pb($n,n'$$\gamma)$        & 14.17(15)      &                               & 0.422(4)                                                                              \\
572.3 &  $^{228}$Ac                               &                         & 0.53(2)                 &                                                                                                              \\
574.7 & $^{74}Ge(n,\gamma)$           &                        &                             & 0.091(2)                                                                               \\
 583.1 & $\left\{ \begin{array}{c}^{208}\mbox{Tl} \\ ^{208}\mbox{Pb($n,n'$}\gamma\mbox{)} \end{array} \right\} $       & 71.64(33)       & 49.6(2)                      & 0.256(3)   \\
 595.9 & $\left\{ \begin{array}{c}^{74}\mbox{Ge($n,n'$}\gamma\mbox{)} \\ ^{73}\mbox{Ge(n,}\gamma\mbox{)} \end{array} \right\} $       & 59.90(30)      &                               & 1.869(7)   \\
608.3 & $^{73}$Ge(n,$\gamma$)            &                          &                              & 0.333(3)                                                                              \\
609.2 & $^{214}$Bi                                & 60.11(30)       &                              &                                                                                                              \\
629.6 & $^{72}$Ge($n,n'$$\gamma$)         &                          &                              & 0.078(2)                                                                             \\
648.2 & $^{115}$In(n,$\gamma$)            &                          &                              & 0.025(1)                                                                              \\
657.2 & $^{206}$Pb($n,n'$$\gamma$)      &                           &                              &  0.047( 1)                                                                              \\
662.0 & $^{137}$Cs                              & 9.04(12)          &                               &                                                                                                              \\
663.8 & $^{206}$Pb($n,n'$$\gamma$)       &                          &                              &0.069(1)                                                                              \\
669.0 & $^{70}$Ge($n,n'$$\gamma$)         &                          &                              & 0.030(1)                                                                              \\
692.4 & $^{72}$Ge($n,n'$e$^{-}$)              & 87.70(37)       &                              &2.406(8)                                                                               \\
701.0 & $^{74}$Ge($n,n'$$\gamma$)         &                          &                             & 0.082(2)                                                                              \\
708.2 & $^{70}$Ge(n,$\gamma$)           &                           &                              & 0.176(2)                                                                              \\
727.3 & $^{212}$Bi                               & 15.72(16)        & 11.7(1)                &                                                                                                              \\
747.7 & $^{70}$Ge(n,$\gamma$)            &                         &                               & 0.047(1)                                                                              \\
755.3 & $^{228}$Ac                              & 2.19(6)            & 1.52(3)                 &                                                                                                              \\
763.1 & $^{208}$Tl                               &                          &  0.85(3)               &                                                                                                              \\
763.1 & $^{208}$Pb($n,n'$$\gamma$)?   &                         &                              & 0.032(1)                                                                              \\
$\left. \begin{array}{c}766.6 \\ 768.4 \end{array} \right. $ & $\left. \begin{array}{c}^{224m}\mbox{Pa} \\ ^{214}\mbox{Bi} \end{array} \right\} $& 4.85(9)           &      & \\
771.8 & $^{228}$Ac                              & 2.02(6)           & 2.11(4)                 &                                                                                                              \\
782.0 & $^{228}$Ac                              &                          & 0.67(2)                &                                                                                                              \\
785.5 & $^{212}$Bi                               & 4.19(8)           & 1.57(3)                 &                                                                                                              \\
$\left. \begin{array}{c}786.3 \\ 786.8 \end{array} \right. $ & $\left. \begin{array}{c}^{35}\mbox{Cl(n,}\gamma\mbox{)} \\ ^{208}\mbox{Pb($n,n'$}\gamma\mbox{)} \end{array} \right\} $&  &   &0.041(1)     \\
$\left. \begin{array}{c}788.4 \\ 788.7 \end{array} \right. $ & $\left. \begin{array}{c}^{35}\mbox{Cl(n,}\gamma) \\ ^{70}\mbox{Ge(n,}\gamma) \end{array} \right\} $&            & &    0.064(1)   \\
 795.0 & $^{228}$Ac                            & 7.92(11)         & 6.1(1)                    &                                                                                                              \\
 798.0 & $^{208}$Pb($n,n'$$\gamma$)    &                           &                              & 0.023(1)                                                                              \\
 803.1 & $^{206}$Pb($n,n'$$\gamma$)    & 20.90(18)       &                               & 0.850(5)                                                                             \\
806.2 &  $^{214}$Bi                             & 3.03(7)             &                              &                                                                                                              \\
 808.2 &  $^{70}$Ge(n,$\gamma$)        &                           &                               & 0.048(1)                                                                              \\
 818.6 & $^{116m}$In                         &                            &                               & 0.064(1)                                                                              \\
 824.9 &                                                 & 1.03(4)              &                              &                                                                                                                \\
830.4  & $^{228}$Ac                           &                            & 0.70(2)                &                                                                                                              \\
 834.1 & $^{72}$Ge($n,n'$$\gamma$)     & 45.15(26)         &                               &  0.290(3)                                                                              \\
835.6 & $^{228}$Ac                            &                            & 2.30(4)                &                                                                                                              \\
840.4 & $^{228}$Ac                            &                            & 1.19(3)                &                                                                                                              \\
 843.8 & $^{27}$Al($n,n'$$\gamma$)       & 4.48(8)             &                               & 0.112(2)                                                                              \\
 846.9 & $^{76}$Ge($n,n'$$\gamma$)    &                           &                               &0.062(1)                                                                              \\
$\left. \begin{array}{c}860.4 \\ 860.4  \end{array} \right. $   & $\left. \begin{array}{c}^{208}\mbox{Tl} \\ ^{208}\mbox{Pb($n,n'$}\gamma\mbox{)}  \end{array} \right\} $       & 8.64(12)          & 6.0(1)                       &  0.090(2)  \\
 865.0& Unidentified                          &                           &                              & 0.094(2)                                                                              \\
 867.9 & $^{73}$Ge(n,$\gamma$)      & 4.25(8)              &                                & 0.466(4)                                                                              \\
 881.0 &  $^{206}$Pb($n,n'$$\gamma$) & 2.50(6)              &                               & 0.151(2)                                                                                              \\
 892.9 & $^{212}$Bi                          &                            &  0.42(2)                &                                                                                                              \\
 894.3 & $^{72}$Ge($n,n'$$\gamma$)   &                            &                              & 0.029(1)                                                                              \\
 897.8 & $^{207}$Pb($n,n'$$\gamma$) & 6.28(10)           &                              &0.199(2)                                                                              \\
 904.1 & $^{228}$Ac                         &                            & 0.94(3)                 &                                                                                                              \\
 911.2 & $^{228}$Ac                         & 48.86(27)        & 37.1(2)                  &                                                                                                              \\
 934.1 & $^{214}$Bi                          & 1.99(6)             &                               &                                                                                                              \\
 958.4 & $^{228}$Ac                         &                            & 0.37(2)               &                                                                                                              \\
 960.9 & $^{74}$Ge($n,n'$$\gamma$)   &                            &                             & 0.095(2)                                                                              \\
 964.4 & $^{228}$Ac                         & 11.22(13)         & 6.2(1)                   &                                                                                                              \\
 968.8 & $^{228}$Ac                         & 26.83(20)        &21.6(1)                  &                                                                                                              \\
 981.0 & $^{206,8}$Pb($n,n'$$\gamma$)&                     &                              & 0.035(1)                                                                              \\
 988.4 & $^{228}$Ac                        & 1.95(5)              &  0.20(1)               &                                                                                                              \\
	$\left. \begin{array}{c}993.7 \\ 995.1  \end{array} \right. $   & $\left\{ \begin{array}{c} ^{74}\mbox{Ge($n,n'$}\gamma\mbox{)} \\ ^{206}\mbox{Pb($n,n'$}\gamma\mbox{)}  \end{array} \right\} $ 
     &                    &                               & 0.026(1)                                                                              \\
 999.5 & $^{74}$Ge($n,n'$$\gamma$)         &                     &                               &0.034(1)                                                                              \\
1001.5 & $^{224m}$Pa                          & 8.03(11)     &                               &                                                                                                              \\
 1004.5 & $^{228}$Ac                            &                      & 0.17(1)                &                                                                                                              \\
 1014.5 & $^{27}$Al($n,n'$$\gamma$)        & 7.46(11)     &                              &0.173(2)                                                                              \\
 1033.1 & $^{228}$Ac                            &                      & 0.20(1)               &                                                                                                              \\
$\left. \begin{array}{c}1040.1 \\ 1040.8  \end{array} \right\} $   & $^{70}$Ge($n,n'$$\gamma$)  &16.89(16)     &                                & 0.210(2)                                           \\
 1063.7 & $^{207}$Pb($n,n'$$\gamma$)    & 8.47(11)    &                               &0.145(2)                                                                              \\
1065.0  & $^{228}$Ac                            &                     &  0.47(2)               &                                                                                                              \\
1078.8  & $^{212}$Bi                             & 1.11(4)      &  0.62(2)               &                                                                                                              \\
1093.9 & $^{208}$Tl  sum 511+583  &                      &   0.90(3)               &                                                                                                              \\
$\left. \begin{array}{c}1095 \\ 1095.8 \\ 1096.9  \end{array} \right. $   & $\left\{ \begin{array}{c} ^{207}\mbox{Pb($n,n'$}\gamma\mbox{)} \\ ^{70}\mbox{Ge(n,}\gamma\mbox{)} \\ ^{116m}\mbox{In}  \end{array} \right\} $      &  7.81(11)      &                              & 0.464(4) \\
1101.3 & $^{74}$Ge($n,n'$$\gamma$)     &                       &                                & 0.123(2)                                                                              \\
1105.6 & $^{74}$Ge($n,n'$$\gamma$)     &                       &                                & 0.019(1)                                                                              \\
1110.4 & $^{228}$Ac sum                   &                      & 0.52(2)                 &                                                                                                              \\
 1120.6 & $^{214}$Bi                           & 11.70(13)    &                              &                                                                                                              \\
1122.5 & $^{228}$Ac sum                   &                      &  0.26(1)                &                                                                                                              \\
1126 & $\left\{ \begin{array}{c}^{208}\mbox{Pb($n,n'$}\gamma\mbox{)} \\ ^{72}\mbox{Ge($n,n'$}\gamma\mbox{)}  \end{array} \right\} $&                      &                              & 0.022(1) \\
1131.6 & $^{73}$Ge(n,$\gamma$)?      & 1.34(5)        &                              & 0.034(1)                                                                              \\
1139.4 & $^{70}$Ge(n,$\gamma$)         & 1.34(5)       &                              & 0.077(2)                                                                              \\
1153.5 & $^{228}$Ac                            &                     &  0.15(1)                &                                                                                                              \\
 1155.2 &$^{214}$Bi                             & 1.08(4)       &                               &                                                                                                              \\
$\left. \begin{array}{c}1164.9 \\ 1166.0  \end{array} \right. $   & $\left\{ \begin{array}{c} ^{35}\mbox{Cl($n,n'$}\gamma\mbox{)} \\ ^{72}\mbox{Ge($n,n'$}\gamma\mbox{)}  \end{array} \right\} $ 
& 0.49(3)      &                              & 0.072(1)                                                                              \\
 1173.5 & $^{60}$Co                             & 13.67(14)  &                              &                                                                                                              \\
 1201.2 & p(n,$\gamma$)d DEP         &                     &                              & 0.415(3)                                                                              \\
	 1204.2 & $\left\{ \begin{array}{c} ^{73}\mbox{Ge(n,}\gamma\mbox{)} \\ ^{74}\mbox{Ge($n,n'$}\gamma\mbox{)}  \end{array} \right\} $& 9.39(12)    & &0.163(2) \\
1226.7 & $^{74}$Ge($n,n'$$\gamma$)      &                   &                              & 0.017(1)                                                                              \\
1238.4 & $^{214}$Bi                              & 5.22(9)      &                              &                                                                                                              \\
1246.9 & $^{228}$Ac                              &                   & 0.58(2)                &                                                                                                              \\
1261.0 & $^{74}$Ge($n,n'$$\gamma$)      &                   &                             & 0.019(1)                                                                              \\
 1281.0 & $^{214}$Bi                              & 0.74(3)     &                               &                                                                                                              \\
1286-7 &  $^{228}$Ac  Blend                &                   &  0.14(1)                &                                                                                                              \\
 1293.5 & $^{116m}$In                          & 4.09(8)       &                              & 0.462(4)                                                                             \\
 1298.8 & $^{70}$Ge(n,$\gamma$)          &                    &                              & 0.087(2)                                                                             \\
 1332.5 & $^{74}$Ge($n,n'$$\gamma$)       &                    &                              & 0.018(1)                                                                              \\
 1332.5 & $^{60}$Co                              &12.27(14)  &                              &                                                                                                              \\
$\left. \begin{array}{c} 1344.5 \\ 1345.9 \\ 1347.7  \end{array} \right. $   & $\left\{ \begin{array}{c} ^{74}\mbox{Ge(n,}\gamma\mbox{)} \\ ^{206}\mbox{Pb($n,n'$}\gamma\mbox{)} \\ ^{70}\mbox{Ge(n,}\gamma\mbox{)} \end{array} \right\} $
 & 0.52(3)      &                              & 0.017(1)                                                                              \\
1374.2  & $\left\{ \begin{array}{c}^{228}\mbox{Ac sum} \\ 964+409 \\911+463  \end{array} \right\} $   &         &0.28(1)      &                                    \\
 1378.0 & $^{214}$Bi                             & 3.24(7)       &                             &                                                                                                              \\
 1378.8 & $^{70}$Ge(n,$\gamma$)         &                      &                             &  0.065(1)                                                                               \\
 1393.8 & $^{206}$Pb($n,n'$$\gamma$)    &                      &                             &  0.016(1)                                                                               \\
 1401.5 &$^{214}$Bi                               & 0.58(3)      &                               &                                                                                                              \\
 1408.6 & $^{214}$Bi                             & 1.84(5)        &                             &                                                                                                              \\
 1413.6 & $^{73}$Ge($n,n'$$\gamma$)      &                      &                             & 0.018(1)                                                                              \\
1431.1 &  $^{228}$Ac                            &                      &  0.15(1)              &                                                                                                              \\
 1433.5 & $^{206}$Pb($n,n'$$\gamma$)      &                    &                             & 0.020(1)                                                                              \\
 1436.9 &$^{208}$Pb($n,n'$$\gamma$)       &                     &                             & 0.017(1)                                                                              \\
 1459.2 & $^{228}$Ac                            &                       & 0.80(2)              &                                                                                                              \\
 1461.0 & $^{40}$K                                & 30.18(22)    &                              & 0.066(1)                                                                            \\
 1463.9 & $^{72}$Ge($n,n'$$\gamma$)    &                         &                              & 0.114(2)                                                                              \\
 1466.8 & $^{206}$Pb($n,n'$$\gamma$)   &                        &                              & 0.032(1)                                                                              \\
 1471.6 & $^{73}$Ge(n,$\gamma$)        &                        &                              & 0.047(1)                                                                              \\
 1489.2& $^{74}$Ge($n,n'$$\gamma$)      &                        &                              & 0.025(1)                                                                              \\
1496.2 & $^{228}$Ac                            & 0.73(3)         &  0.85(3)                &                                                                                                              \\
1501.7 & $^{228}$Ac                            &                         & 0.42(2)              &                                                                                                              \\
1508.9 & $^{116m}$In                          &                        &                               &  0.069(1)                                                                              \\
 1508.9 & $^{214}$Bi                            & 1.95(3)         &                              &                                                                                                              \\
1512.7 &$^{212}$Bi                              &                        & 0.38(2)                &                                                                                                              \\
 1538   &$^{214}$Bi                               & 0.54(3)         &                               &                                                                                                              \\
1557.1 &$^{228}$Ac                             &                        & 0.14(1)               &                                                                                                              \\
1580.8 & $^{228}$Ac                            & 0.68(3)         & 0.55(2)                 &                                                                                                              \\
 1588.3 & $^{228}$Ac                           & 6.06(10)      &2.94(5)                  &                                                                                                              \\
   $\left. \begin{array}{c}1592.5 \\ 1592.5 \\  1593.0 \end{array} \right. $   & $\left. \begin{array}{c}^{208}\mbox{Tl DEP} \\ ^{208}\mbox{Pb DEP} \\ ^{207}\mbox{Pb($n,n'$}\gamma\mbox{)} \end{array} \right\} $       & 7.24(11)      & 2.10(4)                & 0.107(2) \\
 1599.3 & $^{214}$Bi                          & 0.38(2)            &                               &                                                                                                              \\
$\left. \begin{array}{c} 1601.1 \\ 1602.0 \end{array} \right. $   & $\left\{ \begin{array}{c} ^{35}\mbox{Cl(n,}\gamma\mbox{)} \\ ^{74}\mbox{Ge($n,n'$}\gamma\mbox{)} \end{array} \right\} $
&                        &                               & 0.018(1)                                                                              \\
 1614.9 &$^{208}$Pb($n,n'$$\gamma$)?  &                       &                              &  0.020(1)                                                                              \\
 1620.5 & $^{212}$Bi                            & 1.81(5)          & 1.32(3)                 &                                                                                                              \\
 1625.0 & $^{228}$Ac                            &                       & 0.34(2)                &                                                                                                              \\
1630.7 & $^{228}$Ac                            & 1.82(5)          & 1.46(3)                &                                                                                                              \\
$\left. \begin{array}{c}1631.5 \\ 1632.0  \end{array} \right. $   & $\left. \begin{array}{c}^{74}\mbox{Ge($n,n'$}\gamma\mbox{)} \\ ^{70}\mbox{Ge(n,}\gamma\mbox{)} \end{array} \right\} $
 &                        &                              & 0.021(1)                                                                              \\
 1634.0& $^{76}$Ge(n,$\gamma$)          &                      &                              & 0.016(1)                                                                              \\
1638.3 & $^{228}$Ac                            & 0.43(3)         & 0.41(2)                &                                                                                                              \\
   $\left. \begin{array}{c}1640.4 \\ 1640  \end{array} \right. $   & $\left. \begin{array}{c}^{208}\mbox{Pb($n,n'$}\gamma\mbox{)} \\ ^{74,76}\mbox{Ge($n,n'$}\gamma\mbox{)} \end{array} \right\} $
   &                         &                               & 0.026(1) \\
 1661.3    &$^{214}$Bi                          & 0.36(2)         &                               &                                                                                                              \\
1666.3 & $^{228}$Ac                            &                         & 0.17(1)               &                                                                                                              \\
 1699.5 & $^{206}$Pb($n,n'$$\gamma$)?    &                       &                               &  0.021(1)                                                                              \\
 1704.5 & $^{206}$Pb($n,n'$$\gamma$)    & 0.72(3)        &                               &0.041(1)                                                                              \\
 1710.9 & $^{72}$Ge($n,n'$$\gamma$)  &                     &                               &  0.327(3)                                                                               \\
 1712.2 & p(n,$\gamma$)d     SEP     &                        &                               & 0.156(2)                                                                              \\
 1725.7 & $^{207}$Pb($n,n'$$\gamma$)    &                        &                               & 0.029(1)                                                                              \\
 1729.6 & $^{214}$Bi                             & 3.59(7)         &                               &                                                                                                              \\
 1764.7 & $^{214}$Bi                            & 11.68(13)     &                              &                                                                                                              \\
1779.0 & $\left. \begin{array}{c}^{27}\mbox{Al(n,}\gamma\mbox{)}^{28}\mbox{Al} \\ ^{28}\mbox{Al}\Rightarrow^{28}\mbox{Si} \end{array} \right\} $& 2.13(6) &   &0.127(2)                                   \\
1806.0 & $^{212}$Bi                             &                        &  0.11(1)                &                                                                                                              \\
1844.5 & $^{206}$Pb($n,n'$$\gamma$)     &                       &                              & 0.044(1)                                                                              \\
1846.9 & $^{214}$Bi                             & 2.26(6)          &                              &                                                                                                              \\
1940.4& $^{74}$Ge($n,n'$$\gamma$)       &                        &                              &0.027(1)                                                                              \\
1951.1 &  $^{35}$Cl(n,$\gamma$)           &                         &                              & 0.032(1)                                                                              \\
1959.3 &$^{35}$Cl(n,$\gamma$)             &                         &                              & 0.023(1)                                                                              \\
   $\left. \begin{array}{c}2092.1 \\ 2092.7  \end{array} \right. $   & $\left. \begin{array}{c}^{206}\mbox{Pb($n,n'$}\gamma\mbox{)} \\ ^{207}\mbox{Pb($n,n'$}\gamma\mbox{)} \end{array} \right\} $
&                        &                              &0.039(1)  \\
2103.8 & $\left\{ \begin{array}{c}^{208}\mbox{Tl SEP} \\ ^{208}\mbox{Pb SEP} \end{array} \right\} $& 5.21(9)          & 2.25(4)                              & 0.090(2)  \\
2112.1 & $^{116m}$In                          &                        &                              &0.061(1)                                                                               \\
 2118.5    &$^{214}$Bi                          & 0.64(3)         &                               &                                                                                                              \\
2204.0 & $^{214}$Bi                             & 3.73(8)         &                              &                                                                                                              \\
2223.3 & p(n,$\gamma$)d                   &                       &                              &5.813(13)                                                                              \\
2390.5& $^{116m}$In                           &                        &                               &0.015(1)                                                                             \\
2448.5 & $^{214}$Bi                             & 0.51(3)         &                               &                                                                                                              \\
2614.5 & $\left\{ \begin{array}{c}^{208}\mbox{Tl} \\ ^{208}\mbox{Pb($n,n'$}\gamma\mbox{)} \end{array} \right\} $& 39.39(25)    & 16.3(1)                              & 0.729(5)  \\
2650.3 & $^{206}$Pb($n,n'$$\gamma$)?    &                        &                               &0.011(1)                                                                              \\
2686 & sum $^{208}$Tl?                     &                         & 0.12(1)                &                                                                                                              \\
2892 & sum $^{208}$Tl                      &                          & 0.08(1)               &                                                                                                              \\
3061.9 & $^{207}$Pb($n,n'$$\gamma$)    &                     &                             & 0.010(1)                                                                              \\
\end{longtable}
\end{center}

\subsection{Neutron Spectra Simulation}
\label{sec:NeutronSimulation}
Fast neutrons (from 100 MeV to 1 GeV or more) tend to produce
additional neutrons through nuclear reactions as they traverse high-Z
material. In particular the flux of neutrons will increase
several-fold while the average neutron energy decreases through these
processes. As a result, fast neutrons will penetrate deep into a
shield producing additional neutrons at lower energies. These low
energy neutrons ($\sim < $ 20 MeV) give rise to a substantial $\gamma$-ray flux
because (n,$n^{'}\gamma$) cross sections are large near 10 MeV, but become
small at higher energies. Hence it is these secondary lower-energy
neutrons that interact with the shield and detector materials to
produce $\gamma$ rays, which can give rise to background in double-beta
decay experiments. To understand the process by which high energy
neutrons influence the low-energy neutron flux and, in turn, the
observed $\gamma$-ray flux, we simulated neutrons impinging on an outer
shield and tracked how their spectrum changed as the particles
traversed the shield. We also simulated the production of
neutron-induced $\gamma$ rays and how the Ge detector responded to them.
Specifically, we performed simulations of several geometries including:
\begin{enumerate}
\item A simulation of the cosmic-ray produced neutrons with energy up to 1 GeV at our lab in Los Alamos and 
 their propagation through a 10-cm Pb shield. The response of the CLOVER detector to $\gamma$ rays produced by neutron interactions in the shield was simulated.
This simulation, compared to our background data, tests the precision to which we can model neutron production, scattering with secondary neutron production and (n,n'$\gamma$) interactions. 
\item A simulation of the neutron flux induced on the CLOVER from the AmBe source 
 (neutron energy $\leq$11.2 MeV) passing through 15 cm of polyethylene before impinging on the 10-cm Pb shield. Since the neutron flux is of low energy, this simulation tests the precision to which we model (n,n'$\gamma$) interactions.
\item A simulation of the neutron flux with neutron energies up to a few GeV expected at 3200 mwe deep  due to cosmic-ray $\mu$ interactions
in the surrounding rock and 30-cm lead shield and the resulting response of the CLOVER to
 the flux of $\gamma$ rays arising from this flux. This simulation permits us to estimate rates in detectors situated in underground laboratories.
\end{enumerate}
The first two of these simulations are to verify the code's predictive power. The third is to aid in understanding
the utility of depth to avoid neutron-induced backgrounds. 

The simulation package GEANT3-GCALOR~\cite{rbr,czt} is described in detail in Ref. ~\cite{meihime}. 
In general, $(n,n')$ reactions leave the target nucleus in a highly excited state which 
subsequently decays via a $\gamma$-ray cascade to the ground state. 
In the simulation, inelastic scattering cross sections for excitation to a given level depends on the
properties of the ground and excited states. These cross sections were calculated using 
in-house-written code based on Hauser-Feshbach~\cite{wha}
 theory modified by Moldauer~\cite{pam}. The validation of the Hauser-Feshbach theory has been the
 subject of several studies~\cite{hvo,ric,apa}.  The simulated $\gamma$-ray flux arises from the relaxation of
 the initial excited-state distribution, which includes a large number
of levels (60 states for $^{208}$Pb$(n,n'\gamma$) reactions, for example). The nuclear levels and their decay channels
were provided by the ENSDF\cite{ENSDF} database through the GEANT package.  Note however, that the simulation
did not predict every possible transition. In particular the important 2041-keV and 3062-keV emissions
from Pb were not part of this simulation. This situation arises because the simulation packages only have $(n,n')$ cross
sections for 
the lowest lying excited states for most nuclei. It is set to zero for most other levels.
The details of this simulation are described in detail in Ref. ~\cite{meihime}. Here we study the effectiveness of the simulation to predict spectra resulting from (n,n'$\gamma$).

\begin{figure}[htb!!!]
\includegraphics[angle=0,width=7.5cm]{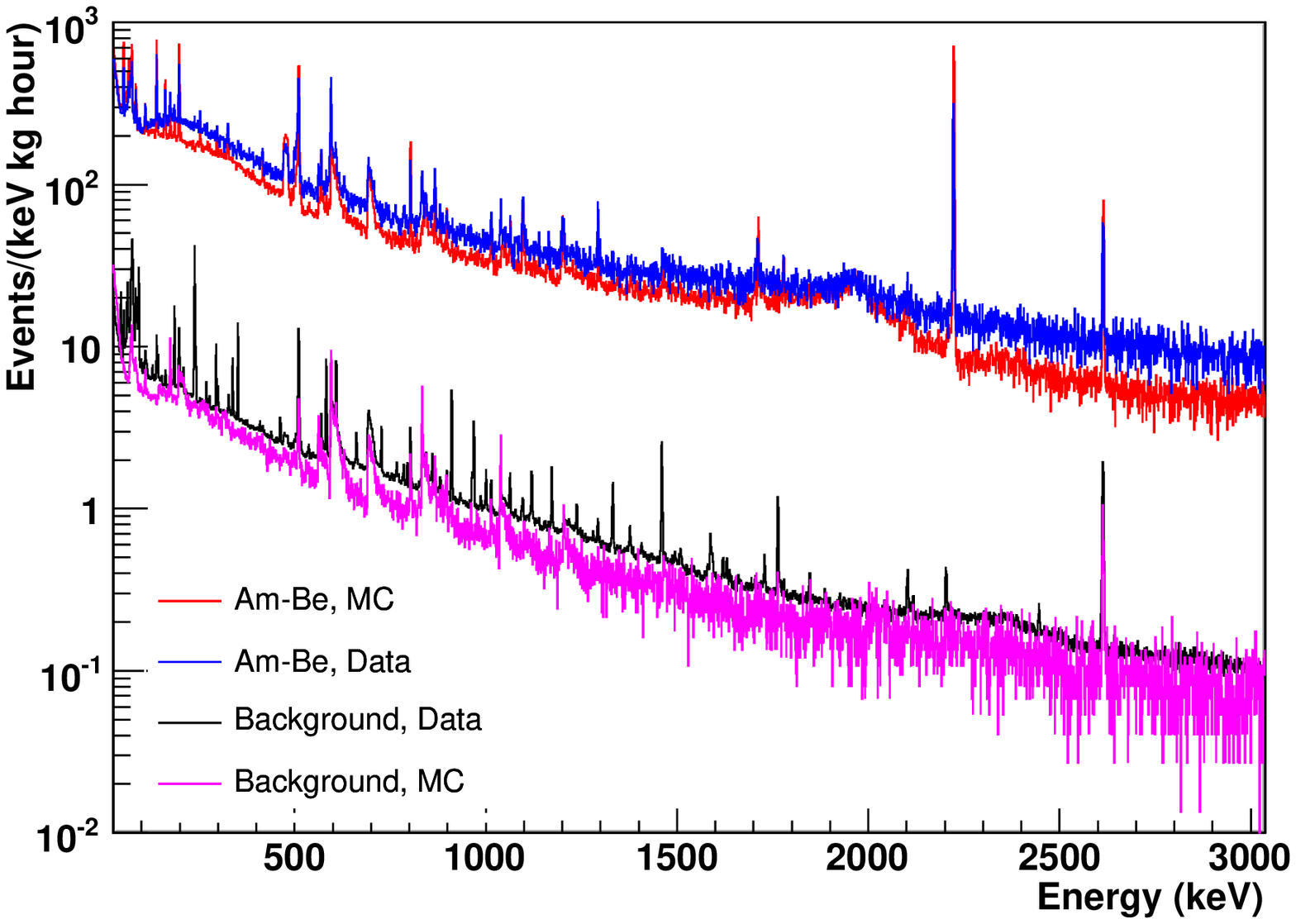}
\includegraphics[angle=0,width=7.5cm]{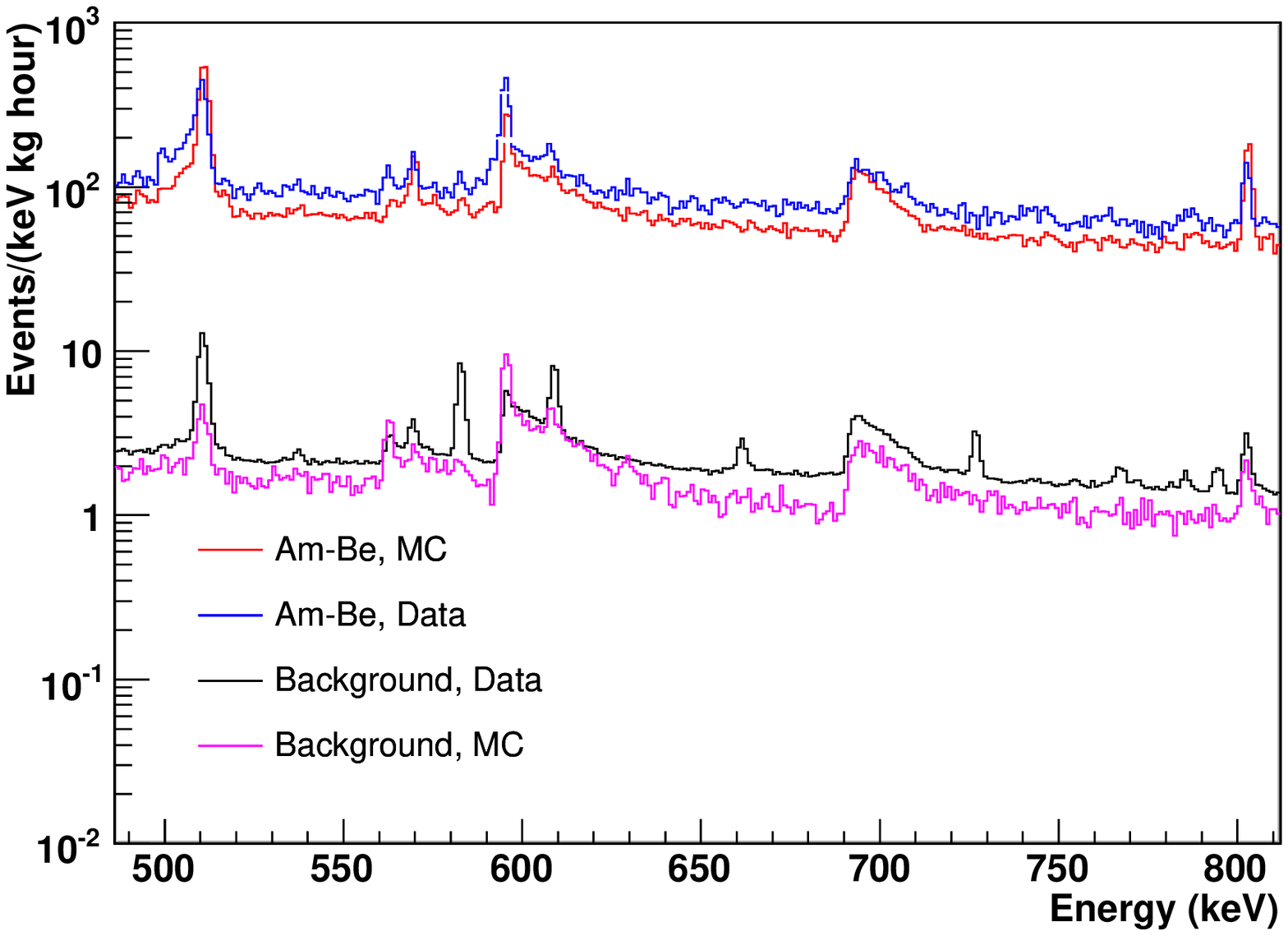}
\caption{\small{Comparison of the measured and simulated AmBe spectra
for the CLOVER detector  {surrounded} by 10 cm lead and 15 cm of moderator. The
upper plot shows the energy range between 10 - 3100 keV. The lower plot
shows the range 470 - 830 KeV where the most significant
$(n,n'\gamma$) lines can be seen. The simulated AmBe neutron
spectrum was normalized to the AmBe source strength for 6.13-h live-time.
The measured total neutron flux in the background spectrum (see Section~\ref{Sec:CosmicFlux}) was used to
normalize the simulated background spectrum. Note, only neutrons as primary particles were simulated
for this comparison and the difference between the spectra is due to the room's
natural radioactivity and non-neutron, $\mu$-induced processes.}}
\label{fig:back}
\end{figure}

The simulation was done by generating neutrons with the appropriate energy spectrum 
outside the lead shield and propagating them through
the shield including secondary interactions that may add to the neutron flux and alter the energy spectrum.  
 Fig.~\ref{fig:back} shows a comparison between the data and the simulations for the CLOVER background run and
AmBe run. 
 Note only neutrons as primary particles were simulated for this comparison and the dominant difference between
the two spectra is due to the room's natural radioactivity and non-neutron $\mu$-induced processes. Here
we excluded those processes from the simulation to emphasize the spectral shape, including
lines, that are a direct result of neutron interactions. 
 The similarity of the spectra in Fig.~\ref{fig:back} indicates that the measured background
spectrum is dominated by neutron-induced reactions. 

The uncertainty in the simulation
is calculated by comparing the well known peaks in Table~\ref{tab:SimulatedLines} which  shows a comparison of 
the simulation to the line production for both background and AmBe runs. 
 The measured neutron produced lines are within about 5\% of the predicted values from simulation, as is the continuum rate in the AmBe data. Therefore the (n,n'$\gamma$) rates are well-simulated for nuclear states with well-defined cross sections. (The continuum for the background data includes processes that were not simulated and hence is not a good measure of the uncertainty.) Because the neutron flux estimates come from these line strengths (See Section~\ref{sec:NeutronFlux}), the uncertainty in the flux cancels in these estimates.
The uncertainty in
the  measured neutron flux and spectrum underground($\approx$35\%) constrains the precision to which such simulations can be verified and is well described in Ref.~\cite{meihime}. 
This 35\% uncertainty due to the flux is much larger than the uncertainty for the $\gamma$-ray line production described above. Therefore, a total uncertainty of 35\% is used for all predictions of line rates underground throughout this paper.

\begin{table}
\small{
\caption{\protect  A comparison of the simulated to measured rates (Background: per hour and AmBe: Hz) for several lines produced by neutron interactions. The 2041-keV and 3062-keV lines are not included in the simulation. } 
\label{tab:SimulatedLines}
\begin{center}
\begin{tabular}{|cc|cc|cc|}
\hline \hline
Process                           &$\gamma$-ray         &\multicolumn{2}{c|}{Background-CLOVER}\\
                                  &Energy               &Simulation    &Measurement\\                                                    
\hline                      
$^{74}$Ge$(n,n'\gamma$)           &596 keV              &56.21                 &59.90(30)\\
$^{74}$Ge$(n,n'\gamma$)           &254 keV              &2.63                &2.76(7)\\
$^{76}$Ge$(n,n'\gamma$)           &2023 keV             &3.2$\times 10^{-7}$ &below sensitivity\\
$^{206}$Pb$(n,n'\gamma$)          &537 keV              &4.82                &5.12(9)\\
$^{207}$Pb$(n,n'\gamma$)          &898 keV              &6.21                &6.28(10)\\
$^{206}$Pb$(n,n'\gamma$)          &1706 keV             &0.69                &0.72(3)\\
$^{206}$Pb$(n,n'\gamma$)          &2041 keV             &none                &not seen\\
Continuum region                  &2000-2100 keV        &110.2               &187.35(19)\\
$^{207}$Pb$(n,n'\gamma$)          &3062 keV             &none                &not seen\\
\hline\hline
Process                           &$\gamma$-ray         &\multicolumn{2}{c|}{AmBe-CLOVER}\\
                                  &Energy               &Simulation         &Measurement\\                                                     
\hline                      
$^{74}$Ge$(n,n'\gamma$)           &596 keV              &1.8                 &1.87\\
$^{74}$Ge$(n,n'\gamma$)           &254 keV              &0.36                &0.41\\
$^{76}$Ge$(n,n'\gamma$)           &2023 keV             &8.5$\times 10^{-4}$ &below sensitivity\\
$^{206}$Pb$(n,n'\gamma$)          &537 keV              &0.15                &0.16\\
$^{207}$Pb$(n,n'\gamma$)          &898 keV              &0.14                &0.20\\
$^{206}$Pb$(n,n'\gamma$)          &1706 keV             &0.04                &0.04\\
$^{206}$Pb$(n,n'\gamma$)          &2041 keV             &none                &not seen\\
Continuum region                  &2000-2100 keV        &7.01               &7.33\\
$^{207}$Pb$(n,n'\gamma$)          &3062 keV             &none                &0.01\\
\hline \hline
\end{tabular}
\end{center}
}
\end{table}

\section{The Neutron Flux}
\label{sec:NeutronFlux}

In this section, we use the data to determine the neutron fluxes we observed during our various experimental configurations.
We then compare our measured cosmic-ray induced flux with that predicted from past measurements and our simulation.

\subsection{Ge$(n,n'\gamma$) Analysis}
Spectral lines that indicate neutron interactions in natural Ge detectors 
have been studied previously. See References~\cite{CHU70, BUN74,LIN90,FEH96}, for example. In particular, the
{\it sawtooth}-shaped peaks due to $^{72,74}Ge(n,n')$ at 693 keV and 596 keV respectively are clear indications
of neutrons and have been used to deduce neutron fluxes~\cite{SKO92}. Operating Ge detectors in a 
low-background configuration, these lines can be used to help 
interpret the background components.  Recent double-beta decay experiments~\cite{igex, HM} have constructed their detectors from
Ge enriched in isotope 76. Although an appreciable amount of $^{74}$Ge remained (14\%), $^{70,72,73}$Ge are depleted.
For such detectors, only lines originating in isotopes 74 and 76 are useful for neutron interaction analysis. As these experiments reach for lower background, neutron-induced backgrounds become a greater concern and the diagnostic tools more important. 

Neutrons from ($\alpha$,n) and fission reactions have an energy spectrum
with an average energy similar to the AmBe spectrum used in this study. Furthermore, the average energy of the AmBe 
neutrons is similar to that of the neutrons
within the hadronic cosmic-ray flux impinging on our surface laboratory  {although the latter extend to much higher energies.} Therefore the Ge-detector signatures indicating the presence 
of neutrons described above will be similar to those arising from neutrons originating from the rock walls of an underground laboratory.
However, low-background experiments that use Ge detectors are typically deep underground and are shielded from environmental
radioactivity by a thick shield. This shield, typically made of Pb, is then usually surrounded by a neutron moderator.
This configuration is effective at greatly reducing the neutron flux originating from ($\alpha$,n) and fission reactions 
in the cavity walls of the underground laboratory. In contrast, although neutrons originating from $\mu$ interactions underground are 
much rarer, they have much higher energy. Therefore these $\mu$-induced neutrons can penetrate the shield more readily and
become a major fraction of the neutrons impinging on the detector.

\subsection{The AmBe Neutron Flux}
\label{Sec:NeutronFlux}
 The estimate of the flux of neutrons with energies greater than 692 keV is given by~\cite{SKO92,phs,rwo}
\begin{equation}
\label{eq:flux1}
\Phi_{n} = \mbox{k}\frac{I}{V},
\end{equation}
where I is the counts s$^{-1}$ under the asymmetric 692-keV peak, V is the volume of the detector in cm$^{3}$ (566 cm$^3$)
and k
is a parameter found by Ref.~\cite{SKO92} to be 900 $\pm$ 150 cm. For the 15-cm moderator data, this formula predicts a neutron flux of 2.3/(cm$^2$ s) whereas our simulation,
 using the known flux of the source,
 predicts 1.8/(cm$^2$ s). This difference (20-30\%) 
is somewhat greater than the 17\% uncertainty claimed by Ref.~\cite{SKO92}.
The geometry for our measurement was complicated and perhaps this added complexity of neutron propagation contributes to the difference.
For the uncertainty associated with the flux of neutrons produced from cosmic ray $\mu$, we use the 35\% value as it is much larger than the value associated with Eqn.~\ref{eq:flux1}.

For the Am-Be neutron source, the rate in the
692-keV peak is 2.406 $\pm$ 0.008 Hz. This results in
 $\Phi_{n}^{ambe}$ = 3.8 $\pm$ 1.1 /(cm$^{2}$ s). This rate is an average over the two moderator configurations. The neutron flux during the 10-cm
 moderator run is estimated to be about a factor 2.3 larger than for the 15-cm moderator run. For the PopTop-AmBe run on Pb for the raw data (in coincidence with the NaI
 detector), the effective flux was 8.6 $\pm$ 2.6/(cm$^{2}$ s) (0.26 $\pm$ 0.08 /(cm$^{2}$ s)).

\subsection{Cosmic-ray Induced Neutron Fluxes}
\label{Sec:CosmicFlux}
In the background spectrum the rate in the 692-keV peak is 87.7 $\pm$ 0.4/hr. 
Using Eqn.~\eqref{eq:flux1} with $I$ = (2.44 $\pm$ 0.06)$\times$10$^{-2}$ Hz for the background spectrum,
one obtains a fast neutron flux of $\Phi_{n}^{back}$ = (3.9 $\pm$ 1.2)$\times$10$^{-2}$ /(cm$^{2}$ s) at the detector in
our surface laboratory.

Ref.~\cite{SKO92} provides a similar formula to estimate the thermal neutron flux, which is accurate to approximately 30\%.
Using the intensity of the
139.68-keV $\gamma$-ray line of $^{75m}$Ge:
\begin{equation}
\label{eq:flux2}
\Phi_{th}(\frac{n}{cm^{2}s}) = \frac{980I_{139.68}}{(\epsilon_{139.68}^{\gamma}+1.6)V},
\end{equation}
with
\begin{equation}
\label{eq:flux3}
\epsilon_{139.68}^{\gamma} \simeq 1 - \frac{1-e^{-V^{1/3}}}{V^{1/3}}
\end{equation}
where $I$ = 47.2 $\pm$ 0.3 /h = 0.013 Hz is the event rate in the peak of 139.68-keV line and V is the volume of the detector in cm$^{3}$.
Using V = 566 cm$^{3}$ we obtain $\Phi_{th}^{back}$ = (9.1 $\pm$ 2.7)$\times$10$^{-3}$ /(cm$^{2}$ s). We also measure
the thermal neutron flux for the Am-Be neutron source, $\Phi_{th}^{ambe}$ = 1.6 $\pm$ 0.5 /(cm$^{2}$ s). 

Thus the total neutron flux incident on the Ge detector measured for the background run is approximately $\Phi_{tot}^{back}$ = $\Phi_n^{back}$ + $\Phi_{th}^{back}$ = (4.8 $\pm$ 0.7) $\times$ 10$^{-2}$ /(cm$^{2}$ s).

\subsection{Neutron Flux as a Function of Depth}
In our basement laboratory, there are 3 primary sources of environmental neutrons. The largest contribution comes from 
the hadronic cosmic ray flux. The next largest arises from $\mu$ interactions in the 77 g/cm$^2$ thick overhead 
concrete layer in the building. Finally there is the negligible contribution from ($\alpha$,n) and fission neutrons
from natural radioactivity in the room. The atmospheric depth at the altitude of our laboratory is 792 g/cm$^2$. Including
the concrete, the depth is 869 g/cm$^2$. Using the analysis of Ziegler~\cite{Zie96,Zie98}, the flux at our lab due to 
the hadronic flux can be estimated to be 3.0 times larger than that at sea level. The flux at sea level has been
measured to be 1.22 $\times$ 10$^{-2}$ /(cm$^{2}$ s)~\cite{pgo} resulting in a flux in our laboratory of 3.7 $\times$ 10$^{-2}$ /(cm$^{2}$ s).
To estimate the additional neutron flux originating from $\mu$ interactions in the concrete above our laboratory, we rely on our simulations
of neutron generation and propagation. The simulation predicts 1.4 $\times$ 10$^{-2}$ /(cm$^{2}$ s) (3.3 $\times$ 10$^{-2}$ /(cm$^{2}$ s))
for the muon-induced (hadronic) neutron flux inside the lead shield for a total simulated neutron
flux of 4.7 $\times$ 10$^{-2}$ /(cm$^{2}$ s) in acceptable agreement with our measurement of (4.8 $\pm$ 2.2) $\times$ 10$^{-2}$ /(cm$^{2}$ s) =
(1.51 $\pm$ 0.69)  $\times$ 10$^{6}$ $\pm$ /(cm$^{2}$ y). The 
success of this simulation lends credence to the neutron flux estimate in the following sections.

 

The neutron flux onto the detector will be increased due to the neutron interactions with shield materials and
neutron back-scattering from the cavity walls. For
example, our simulations show that the fast neutron flux will increase by a factor of $\approx$ 10 by traversing a 30-cm lead layer.
Also, neutrons will backscatter from the cavity walls and reflect back toward the experimental apparatus, effectively
increasing the impinging neutron flux by a factor of 2-3 depending on the specific geometries of the detector and experimental hall. 
Therefore,
it is important to account for these effects when estimating the neutron flux at the detector.

\begin{figure}[htb!!!]
\includegraphics[angle=0,width=7.5cm]{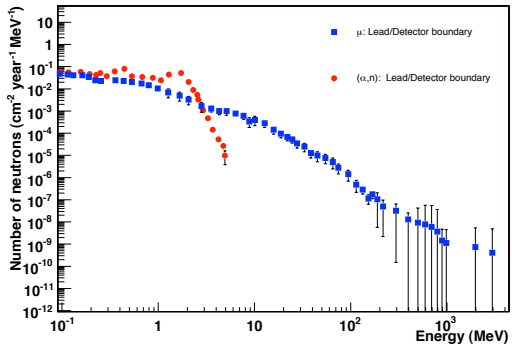}
\caption{\small{
The effective neutron flux onto the simulated
detector described in the text at a depth of 3200 mwe. Shown are the
neutron flux from two sources: (1)  the effective neutron flux induced
by muons that transverse the surrounding rock and shielding materials
assuming a 90\% muon-veto efficiency and (2) the neutron flux from
($\alpha$,n) reactions in the rock.}}
\label{fig:boundary}
\end{figure}

Muon-induced neutron production in different shielding materials and in the detector  itself was also studied in 
Ref.~\cite{meihime}. For example, with 30 cm of lead surrounding a CLOVER-style detector at 3200 mwe, the total muon-induced neutron 
flux impinging on the  detector was calculated to be
 (8.6 $\pm$ 4.0) $\times$ 10$^{-8}$ /(cm$^{2}$ s) = 2.7 $\pm$ 1.2/(cm$^{2}$ y).
 Some of the interactions resulting from these neutrons would be eliminated
by a $\mu$ veto. Assuming a veto efficiency of 90\% for muons traversing this lead shield, the effective
neutron flux is estimated to be (2.0 $\pm$ 0.9) $\times$ 10$^{-8}$ /(cm$^{2}$ s) = 0.63 $\pm$ 0.29 /(cm$^{2}$ y). 
 The energy spectrum
of neutrons at the lead/detector boundary at 3200 mwe is shown in Fig.~\ref{fig:boundary} and has an average value of 45 MeV.

The average energy of the $\mu$-induced neutrons  is 100-200 MeV and much higher than 
that of ($\alpha$,n) neutrons ($\approx$ 5 MeV). The simulated flux of the $\mu$-induced 
neutrons ((2 $\pm$ 0.9) $\times$ 10$^{-8}$ /cm$^{2}$ 
s = 0.63 $\pm$ 0.29/cm$^{2}$ y) inside the detector shield at a depth of 3200 mwe is a factor 2.4 
greater than the simulated ($\alpha$,n) flux surviving the shield 
((0.85 $\pm$ 0.39)$\times$ 10$^{-8}$ /cm$^{2}$ s = 0.27 $\pm$ 0.12/cm$^{2}$ y). The average
energy of these ($\alpha$,n) originating neutrons is 3-5 MeV at the detector surface.

With this estimate of the neutron flux at depth and with our measurements of the Pb and Ge neutron-induced
detector response, we can proceed to estimate these processes in underground Ge-detector experiments.
There are effects in addition to the incident flux, however, that must be taken into account when extrapolating our surface
laboratory results to different geometries and locations. 

\begin{enumerate}
\item As the thickness of the Pb shield increases, additional secondary
neutrons will be generated. Our simulation predicts that a factor $k_{shield}$ = 2.16 more neutrons will be produced by a 30-cm thick shield
as compared to a 10-cm thick shield. 
\item As the energy of the neutrons increases, the number of multiply scattered neutrons increases and therefore the number of interactions that might produce a $\gamma$ ray increases.
For the average energy of neutrons at our surface laboratory (at 3200 mwe),
the average scattering length is $\lambda_{L}$=7.1 cm ($\lambda_{UG}$ = 12.5 cm).
\item Also as the energy increases, the number of states that can be excited in the target nucleus increases.
In the shield at our surface laboratory (at 3200 m.w.e), the average neutron
energy is 6.5 MeV (45 MeV). 

\end{enumerate}

\begin{figure}[htb!!!]
\includegraphics[angle=0,width=7.5cm]{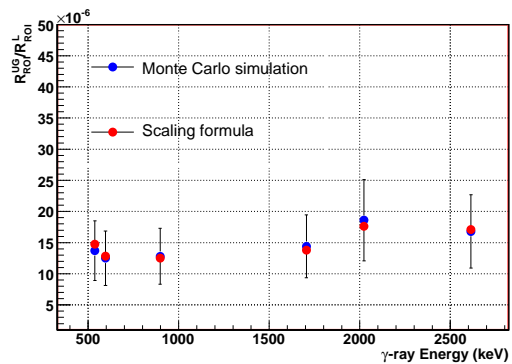}
\caption{\small{
 The comparison between the Monte Carlo simulation of a detector as described in the text and the scaling formula for several excitation lines. The 35\% uncertainties shown in this figure arise from the cross section uncertainty and the statistical uncertainty of determining the peak counts in the simulated spectra. The latter dominates.
}}
\label{fig:scaling}
\end{figure}

All of these factors can be incorporated into a scaling formula derived from our simulation. The rate ($R_{ROI}^{UG}$) of background near the region of interest (ROI)
in an underground laboratory can be related to that measured in our surface laboratory ($R_{ROI}^{L}$) as

\begin{equation}
\label{eqn:ScalingRule}
R_{ROI}^{UG} =    (1+\frac{\lambda_{UG}}{\lambda_L})k_{shield} ((\frac{E_n^{UG}-E_x}{E_n^L-E_x})^{0.8})\frac{\Phi_n^{UG}}{\Phi_n^{L}} R_{ROI}^{L},
\end{equation}

\noindent where $\Phi_n^{L}$ ($\Phi_n^{UG}$) is the neutron flux in our surface laboratory (at 3200 mwe), $E_n$ is the neutron energy and $E_{x}$
is the excitation energy for a typical level. This formula reproduces our simulated results well and the uncertainty of its use is dominated by the precision of simulation. Using the 2.6-MeV level
 in $^{208}$Pb 
as an example, $R_{ROI}^{UG} \sim$ 1.7 $\times$ 10$^{-5} R_{ROI}^{L}$. Fig.~\ref{fig:scaling} shows a comparison between the Monte Carlo simulation and
the scaling formula Eqn.~\ref{eqn:ScalingRule} for several lines. 

\section{Analysis}
\label{sec:anal}

\subsection{Pb$(n,n'\gamma$) analysis}
If fast neutrons are present, then one will also see $\gamma$-ray lines from Pb$(n,n')$ interactions. In very low background
configurations, $\gamma$ rays from neutron-induced excitations in $^{208}$Pb and $^{207}$Pb can be masked by
or confused for decays of $^{208}$Tl and $^{207}$Bi respectively. Therefore it is the stronger transitions in $^{206}$Pb 
(537.5, 1704.5 keV) that
are most useful for determining if these processes are taking place. In $^{207}$Pb the relative strength of the 898-keV transition, compared to the 570- and 1064-keV transitions, is much stronger when it originates from $^{207}$Pb(n,n'$\gamma$) as opposed to $^{207}$Bi $\beta$ decay to $^{207}$Pb.
Therefore this line
can also be used as a tell-tale signature of neutron interactions.

Our data show indications of $^{206,207,208}$Pb$(n,n'\gamma$). As noted earlier, the 2614-keV $\gamma$ ray from $^{208}$Pb can originate from $^{208}$Tl decay or from $^{208}$Pb(n,n'$\gamma$). The 692-keV peak arises only from neutron interactions on $^{72}$Ge. Since the Pb shielding was similar in both the background and AmBe
runs, we can compare the ratio of the rate in the 
2614-keV peak to that in the $^{72}$Ge$(n,n'\gamma$) 692-keV peak in the two data sets to deduce the fraction of the 2614-keV in the background run that can be attributed to neutron interactions.
This ratio in the Am-Be spectrum is 0.30 and that in the 
background spectrum is 0.45, and therefore, we conclude that $\approx$ 67\% of the strength in the background run is due to neutron reactions
and the remainder is due to $^{208}$Tl decay. Clearly, in our surface laboratory, environmental neutrons are
a significant contributor to the observed signal.

\subsection{The Special Cases of the 2023-keV, 2041-keV and 3062-keV $\gamma$ rays}
The 2023-keV level in $^{76}$Ge can be excited by neutrons. The simulation predicts that this line
is too weak to be seen in the CLOVER AmBe data, but the CLOVER is built of natural Ge. In the enriched detectors 
planned by future double-beta decay experiments, the fraction of isotope 76 is much larger and this line would be enhanced.
Still our simulation (Table~\ref{tab:KeyCountRates}) predicts it would be a very small peak.

The 3714-keV level in $^{206}$Pb can emit a 2041-keV $\gamma$ ray. We only observed a candidate $\gamma$-ray peak in the 
coincidence  data (Ge detector event in coincidence with a 4.4-MeV $\gamma$ ray in the NaI detector) with the AmBe source radiating the Pb shield around the PopTop detector. The magnitude of this peak if it exists is small and not convincing. We use this data set to place a limit on the production rate of this line as it results in the most conservative limit.

\begin{figure}[htb!!!]
\includegraphics[angle=0,width=7.5cm]{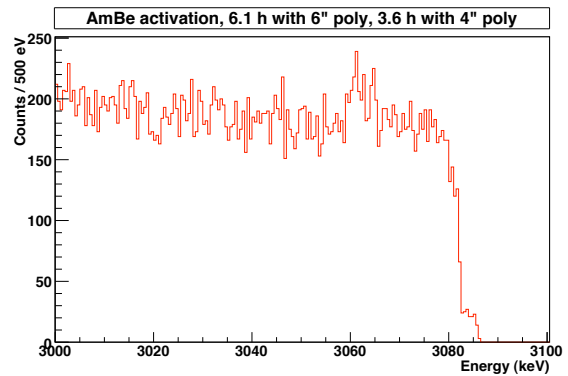}
\caption{\small {The energy spectrum near 3000 keV showing the $^{207}$Pb$(n,n'\gamma$) 3062-keV $\gamma$-ray line
in the AmBe spectrum with the CLOVER.}}
\label{fig:E3062}
\end{figure}

In the AmBe-irradiated CLOVER and the non-coincidence PopTop spectra, we observed a 3062-keV $\gamma$ ray that we
assign to a transition from the 
3633-keV level in $^{207}$Pb. (See Fig. \ref{fig:E3062}.) This line is only present when Pb surrounds the detector: It is absent when Cu forms
the shield. The statistical sensitivity was too weak in the PopTop coincidence spectrum
to observe this weak line.
In the CLOVER AmBe spectrum, the rate of this line
is 5.3 $\times$ 10$^{-3}$ that of 596-keV $^{74}$Ge peak rate and in the raw AmBe PopTop with Pb spectrum 
the ratio is  4.5 $\times$ 10$^{-3}$.  From these data we can estimate an approximate rate that these dangerous 
backgrounds would be produced for a given neutron flux. In our surface laboratory, the CLOVER background rate for
the $^{74}$Ge 596-keV peak was 59.9 events/hr. This leads to a predicted rate of  0.3 events/hr in
the 3062-keV peak. Note that our data indicate that any peak at 3062 keV is statistically weak ($\leq$ 0.2 events/hr)  but reasonably consistent with this prediction. Note, in the AmBe-CLOVER runs, polyethylene blocks were used to increase the flux of thermal neutrons. It appears that these blocks contain some Cl and therefore we see indications of Cl(n,$\gamma$) lines. Even though $^{35}$Cl has a neutron capture line at 3062-keV, we do not assign the observed line in our data to that process. Because the $^{35}$Cl(n,$\gamma$) line at 2863 keV is not observed and because the line at 1959 keV is weak, we conclude that assigning this line to Cl would be inconsistent with the predicted line ratios for neutron capture. However, a concern regarding our assignment of the 3062-keV line to $^{207}$Pb is that one also expects a 2737-keV emission from the same 3633-keV level. This companion $\gamma$ ray is not observed in our data and we plan future measurements dedicated to measuring the neutron-induced relative intensities of these two lines. If we assume that the entire rate (0.01 Hz) of the 3062-keV line is due to $(n,n'\gamma$), we can make a crude estimate of the cross section by scaling to the rate in the 2614-keV line. The cross section for the 2614-keV (n,n'$\gamma$) is 2.1 b $\pm$ 10\% \cite{hvo}. Using this cross section, the relative rates in the two peaks, the different isotopic ratios of $^{208}$Pb and $^{207}$Pb, and the different $\gamma$-ray detection efficiencies, the average cross section for $^{207}$Pb(4.5-MeV n, $n'$ 3062-keV $\gamma$ ray) is estimated to be 75 mb. The uncertainty is estimated to be about 20\% or $\sim$ 15 mb.

 From measurements with the CLOVER and a $^{56}$Co source, which has $\gamma$-ray energies near 3100 keV,
we expect 0.13 DEP events per full-energy $\gamma$-ray event. Therefore, in our surface lab, we expect 0.03 events/hr
in the DEP at  2039 keV due to $^{207}$Pb$(n,n'\gamma$). This is well below our continuum rate of  2.5 events/keV-hr or 10/hr
in an energy window corresponding to a 4-keV wide peak. 

\begin{table*}
\small{
\caption{\protect The raw count rates for select processes normalized to neutron flux of $1/(cm^2 y)$.  When extrapolating to neutron energies distant from
that near the measurements, the uncertainty (35\%) associated with the extrapolation must be included. See text for a discussion of the uncertainty estimates in this table especially with respect to the average. }
\label{tab:SpecialLines}
\begin{center}
\begin{tabular}{ccccc}
\hline \hline
Process                                                           &Rate          &Rate              &Rate         &Rate\\
                                                                       &AmBe-CLOVER   &AmBe-PopTop  &Background   &Average\\
                                                                          &\multicolumn{4}{c}      {($\frac{events}{(t y)}$)}\\
\hline
$^{206}$Pb$(n,n' 537-keV\gamma$)        &13.9$\pm$4.9    &20.5$\pm$7.2         &12.2$\pm$4.3    &15.5$\pm$5.4\\
 $^{74}$Ge$(n,n' 596-keV\gamma$)         &164$\pm$57      &unresolved\footnote{In the PopTop data, the 596-keV line was not resolved from nearby lines.}      &142$\pm$50     &153$\pm$54\\
$^{207}$Pb$(n,n' 898-keV\gamma$)        &17.5$\pm$6.1   &21.2$\pm$7.4          &14.9$\pm$5.2   &17.9$\pm$6.3\\
$^{206}$Pb$(n,n' 1705-keV\gamma$)       &3.6$\pm$1.3     &3.2$\pm$1.1            &1.7$\pm$0.6      &2.8$\pm$1.0\\
$^{206}$Pb$(n,n' 2041-keV\gamma$)       &$<$6.3               &$<0.5$                      &$<$1.7                &$<5.0$\footnote{The 2041 line was not observed in any of
our spectra, however, a weak peak-like feature was present in the AmBe-PopTop coincidence data. We used the upper limit for the rate in that peak as the "average" as we considered this to be most conservative.}\\
$^{207}$Pb$(n,n' 3062-keV\gamma$)       &0.88$\pm$0.31 &0.6$\pm$0.21        &$<$1.0                &0.7$\pm$0.2\\
\hline
                                                                          &\multicolumn{4}{c}{($\frac{events}{(keV t y)}$)}\\
Continuum Rate from $Pb,Ge(n,n'\gamma)$ &2.6$\pm$0.9   &2.0$\pm$0.7       &2.5$\pm$0.9  &2.4$\pm$0.8\\
\hline \hline
\end{tabular}
\end{center}
}
\end{table*}

Since our simulation does not predict all these lines,
we summarize the measured rates normalized to the neutron flux in Table \ref{tab:SpecialLines}  to provide simple
scaling to different experimental configurations. The uncertainties in Table  \ref{tab:SpecialLines} are estimates based on a minor contribution
of the statistical uncertainty in the peak strengths and a major contribution resulting from the $\approx$ 35\% uncertainty in the neutron
flux determination as described in Section \ref{Sec:NeutronFlux}. Because the uncertainty is mostly systematic, there is a good possibility that the total
uncertainties for each individual measurement are correlated. Therefore, to estimate the average values in this Table, we took a straight average of the
individual values and then assigned an uncertainty equal to the largest fractional value. This procedure, although not rigorous, is more conservative than 
a weighted average. In addition, some peaks were not observed in all spectra. The upper limits on the strength of these peaks were estimated from the 
rates of weakest peaks observed near the associated energy region in the spectrum. Such peaks are considered to represent the level of sensitivity
of our peak detection procedures. The 2041-keV line is a special case. We quote an upper limit based on the only spectrum that indicated a possible peak. 

These measurements were done for a 
CLOVER detector inside a 10-cm Pb shield. The relative energy-dependent efficiency ($\epsilon_{rel}$) for a 
full-energy peak in the CLOVER can be approximated by,

\begin{equation}
\epsilon_{rel} = 0.15 + 0.93e^{\frac{-(E_{\gamma} - 148)}{766}},
\end{equation}

\noindent where $E_{\gamma}$ is the $\gamma$-ray energy in keV. This expression is normalized to 1.0 at 209 keV and is estimated to
have an accuracy of about 20\% near 200 keV improving to about 10\% at 2600 keV. The quoted
relative efficiency for each of the 4 individual CLOVER detectors is 26\% 
at 1.33 MeV as quoted by the manufacturer. Table \ref{tab:SpecialLines} does {\em not} incorporate this efficiency correction, therefore the table presents the measured count rates with a minimum of assumptions.
The thickness of Pb is large compared to the mean free path of the 
$\gamma$ rays of interest, therefore, the scaling should hold for other thick-shield configurations. Even so,
the rates will be geometry-dependent so these results can only be considered guides when applied
to other experimental designs. The rate of these excitations also depends on neutron energy. For the background
run (AmBe run) the average neutron energy is $\approx$ 6.5 MeV ($\approx$ 5.5 MeV).  Our simulations predict that the rate of these excitations scales as energy to the 0.81 power.

\section{Discussion}
\label{sec:disc}
\subsection{A Model of the CLOVER Background}
We can use these experimental results to create a background model for our surface lab
 and deduce the contribution to the continuum near 2039 keV due to (n,n'$\gamma$) reactions. We then use simulation of high-energy neutron production and propagation to
extrapolate this model to better understand experiments done at depth. The measured rate for 
the continuum near 2039 keV was 14.8 events/(keV kg d). For the Th-wire
data, this continuum rate was 0.10 events/(keV s) (2900 events/(keV kg d)) and for the AmBe data it was 0.09 events/(keV s)
(2600 events/(keV kg d)). To determine the neutron-induced continuum rates in the AmBe data, 
however, we have to correct for the contribution from the tail of two high-energy
$\gamma$ rays that are not part of the neutron-induced spectrum in the background. These are the $\gamma$ rays
 from the 2223-keV p$(n,\gamma$)d
and the 4.4-MeV $\gamma$ rays originating from the ($\alpha$,n) reaction of the AmBe source itself. Although
only $\approx$ 10\% of these AmBe $\gamma$ rays penetrate the 5-cm Pb shield, there is still a significant flux.

A simple
simulation of the detector response to 2223-keV $\gamma$ rays can easily determine ratio of the rate in the 2039-keV
region to that in the full-energy peak. Simulation indicates that this ratio is 5.2 $\times$ 10$^{-2}$ /keV. Since the full-energy peak
count rate is 5.813 Hz, we find this contribution to the continuum is 0.03 events/(keV s). For the 4.4-MeV $\gamma$ ray from the
source itself, simulation must determine an absolute rate in the continuum because the high-energy threshold
prevented the observation of the full-energy peak or its escape peaks. The simulation predicts 0.03 events/(keV s). Subtracting these two 
contributions from the continuum rate for the AmBe source near 2039 keV results in a final value of 0.03 events/(keV s) or
860 events/(keV kg d).

Our background measurements were done without a cosmic ray anti-coincidence system.
From auxiliary measurements with a scintillator in coincidence with the CLOVER and a similar
shielding geometry, we measured the rate of $\mu$ passing
through the detector. In the continuum near the 2039-keV region, the rate is 5.4 events/(keV kg d).

From the Th-wire source data, we measure the ratio of the rate in the continuum near 2039 keV to 
that in the 2614 keV (16.3 Hz) peak to
be 6 $\times$ 10$^{-3}$/keV. Of the 2614-keV peak rate in the background data, $\approx$ 33\% is due to $^{208}$Tl decay. 
Scaling from the 2614-keV peak in the background data, the count rate near the 2039-keV region  due to the Compton tail
of the $^{208}$Tl 2614-keV peak is 0.7 events/(keV kg d).

The remainder of the 2614-keV peak is due to neutron-induced processes. The contribution due to neutrons 
 can be estimated from the AmBe data.
For the AmBe data, the ratio of the rate in the continuum near the 2039-keV region (0.03 events(keV s)) to that for 
the 596-keV $^{74}$Ge$(n,n'\gamma$) peak (1.87 Hz) is 1.6 $\times$ 10$^{-2}$/keV. Scaling from the $^{74}$Ge
peak rate in the background data (59.9/h) indicates a rate of 7.8 events/(keV kg d) in the continuum near 2039 keV. That is, 
53\% of the events in that region are due to neutrons. One can do a similar scaling from the 692-keV $^{72}$Ge rates. Here
the ratio is 1.3 $\times$ 10$^{-2}$/keV and the continuum rate is 8.8 events/(keV kg d). 
We use the average of the Ge values as our estimate (8.3 events/(keV kg d) = 3030 events/(keV kg y))
 for the neutron induced contribution to the continuum rate.
Table \ref{tab:ROIcontent} summarizes the deduced contributions to the spectrum in the 2039-keV region
 in the CLOVER background spectrum and the following section discusses how these data are used along with simulation to estimate rates in experimental apparatus underground.

\begin{table}
\small{
\caption{\protect A summary of the count rate in the CLOVER background data in the energy region near 2039 keV  
based on the model deduced for the surface lab described in the text. The precision
of the neutron-induced and muon-induced spectra simulations (Section~\ref{sec:NeutronSimulation} and Ref~\cite{meihime}) is estimated to be about 35\%. We take this to be a conservative estimate for the 
uncertainties associated with this Table.}
\label{tab:ROIcontent}
\begin{center}
\begin{tabular}{cc}
\hline \hline
Process                       &CLOVER Event Rate\\
                              &Surface Lab\\
                              &events/(keV kg d)\\
\hline
neutron-induced               &8.3$\pm$2.9\\
$^{208}$Tl Compton scattering &0.7$\pm$0.3\\
high energy $\mu$ continuum   &5.4$\pm$1.9\\
\hline
Total from model              &14.4$\pm$5.0\\
\hline
Measured Rate                 &14.8$\pm$0.2\\
\hline \hline
\end{tabular}
\end{center}
}
\end{table}

\subsection{Solving the Problem with Overburden}
The primary purpose of this study is to better understand the impact of neutrons on the background for future double-beta decay
experiments. 
In this subsection, we use neutron fluxes from our simulations of the surface laboratory,
 measurements with the AmBe source, and simulations of the neutron flux in an underground 
 laboratory to estimate the contribution of neutron-induced 
backgrounds underground. In the following subsection, we examine data from previous underground experiments.

The simulation of neutron processes in the 10-cm Pb shield and Ge comprising the CLOVER detector
at the altitude of our laboratory predicts about 1594$\pm$558 events/(keV kg y) between 2000 and 2100 keV
due to lead excitation and about 1337$\pm$468 events/(keV kg y) in this energy region due to germanium excitation.
Our measured value for the neutron-induced events is 3030$\pm$1061 events/(keV kg y) to be compared with this predicted value of 2931$\pm$1026 events/(keV kg y).

 The simulation of the CLOVER within a 30-cm lead shield at a depth of 3200 mwe, predicts
  about 0.019$\pm$0.007 events/(keV kg y) contributed from lead excitation and about 0.016$\pm$0.006  events/(keV kg y) contributed from 
germanium excitation for a total of 0.035$\pm$0.012 events/(keV kg y). 
One can also just scale our surface-laboratory measurement of the neutron-induced rate near 2039 keV by
the factor derived from Eqn. \ref{eqn:ScalingRule} above. This results in 0.05$\pm$0.02 events/(keV kg y).
 For a detector like the CLOVER, analysis based on pulse shape discrimination (PSD),
and the response of individual segments or crystals can help reduce background based on its multiple-site energy deposit
nature. These backgrounds can then be distinguished from the single-site energy deposit character of double-beta decay.
We have measured the background reduction factor via these techniques to be $\approx$ 5.9 for the CLOVER~\cite{ELL05}.

Reference~\cite{meihime} provides a {\it quick reference} formula to estimate the neutron flux as a function of depth.
 The $\mu$ flux and its associated activity is reduced by $\approx$ 10 for each 1500 m.w.e of added depth. Future
 double-beta decay experiments hope to reach backgrounds near 0.25 events/(keV t y). Our estimate of 
 the rate at 3200 mwe is 35-50 events/(keV t y), which is a factor of 150 above the goal.
 Hence, greater depths would be desirable.

\begin{table*}
\small{
\caption{\protect A summary of the key count rates arising from neutron interactions
in the CLOVER background data in the energy region near 2039 keV
as predicted by our analysis for three representative depths. The shield thickness is taken to be 30 cm and a veto system with an assumed efficiency of 90\%
is included.
Except for the 2023-keV line, we used the scaling of Eqn.~\ref{eqn:ScalingRule} to scale our CLOVER background measurements to the 3200 mwe depth and
then used the muon fluxes at WIPP, Gran Sasso, and SNOLAB\cite{meihime} to scale to the other depths. The Ge rates are 
also scaled for an enriched detector (86\% isotope 76, 14\% isotope 74). The scalings require the results from the simulations.
The uncertainty  
 is dominated by the simulated flux uncertainty and is estimated to be 35\%. Since we did not observe the 2023-keV line, we used simulation to
predict the rate. We used the measured upper limit for the 2041-keV line.
For comparison, the results of Ref.~\cite{Kla04} is shown. Reference~\cite{Kla04} claims a result for zero-neutrino double-beta
 decay in an experiment performed at 3200 mwe. We entered the claimed
event rate for that process in the same row as the 2041-keV line for comparison. The rate limits for the other lines assigned to
Ref.\cite{Kla04} result from our estimates based on the figures in their papers and does not come directly from their papers.}
\label{tab:KeyCountRates}
\begin{center}
\begin{tabular}{ccccc}
\hline \hline
Process             &1600 mwe &3200 mwe    &6000 mwe   &Ref. ~\cite{Kla04}\\
\hline
$^{74}$Ge 596 keV   &19400$\pm$6800/(t y)      &1130$\pm$400/(t y)     &15$\pm$5/(t y)               &$<$800/(t y)\\
$^{76}$Ge 2023 keV  &30$\pm$10/(t y)                 &2$\pm$1/(t y)                &0.02$\pm$0.01/(t y)      &300/(t y)\\
$^{206}$Pb 537 keV  &4400$\pm$1500/(t y)        &250$\pm$88/(t y)         &3.4$\pm$1.2/(t y)         &\\
$^{207}$Pb 898 keV  &5300$\pm$1900/(t y)        &310$\pm$110/(t y)         &4.2$\pm$1.5/(t y)           &\\
$^{206}$Pb 1705 keV &610$\pm$210/(t y)           &36$\pm$13/(t y)           &0.5$\pm$0.2/(t y)            &\\
$^{206}$Pb 2041 keV &$<$1300$\pm$450/(t y)  &$<$74$\pm$26/(t y)   &$<$1.0$\pm$0.3/(t y)     &400/(t y)\\
$^{207}$Pb 3062 keV &145$\pm$51/(t y)             &8.4$\pm$2.9/(t y)         &0.1$\pm$0.03/(t y)         &$<$71/(t y)\\
continuum           &880$\pm$310/(keV t y)             &50$\pm$17/(keV t y)   &0.7$\pm$0.24/(keV t y) &110/(keV t y)\\
\hline \hline
\end{tabular}
\end{center}
}
\end{table*}
 
\subsection{Discussion of Previous Underground Experiments}
Previous Ge-based double-beta decay experiments conducted deep underground~\cite{igex,HM} set the standard for low levels
of background. The future proposals however~\cite{Gerda, MJ} hope to build experiments with much lower backgrounds. In this subsection,
we estimate how large the neutron contribution was to the previous efforts and future designs.
Using the scaling summarized in Eqn.~\ref{eqn:ScalingRule}, we can compare the expectations of our simulated underground 
apparatus with previously published results. Table~\ref{tab:KeyCountRates} shows this comparison. This table also presents a summary
of how the rates would be affected by a change in depth only. The IGEX collaboration~\cite{igex} has not published its data in sufficient
detail to do a similar comparison. Other underground Ge detector experiments do not have the required sensitivity.

The Heidelberg-Moscow experiment~\cite{HM} is a critical case study for such backgrounds and it was operated at a depth of 3200 mwe.
 One is clearly led to consider if the 3062-keV $\gamma$ ray
can explain the signal reported in Ref.~\cite{cdo, Kla04}. Figure 36 in Ref.~\cite{Kla04}, shows that no more than a few counts can be
assigned to a 3062-keV $\gamma$ ray. If the 23 counts assigned to double-beta decay were actually a DEP from this $\gamma$ ray,
the one would expect 175 counts or so in the 3062-keV peak. Therefore, it is difficult to explain the claimed peak by this mechanism. It is also clear that the
predicted rate of the 2041-keV $\gamma$ ray is too low to explain their data. The data from Ref.~\cite{Kla04} show lines at 570 and 1064 keV and the authors assigned these lines to $^{207}$Bi present in the Cu. However, the spectra displayed in Fig. 13 of that paper shows that only
detectors surrounded by Pb indicate the 570-keV line. Since there is no evidence for the 898-keV line in the data, we agree with
the $^{207}$Bi assignment, however, we hypothesize that it must reside in the Pb and not the Cu. Perhaps 
this contamination is cosmogenically produced
in Pb when it resides on the surface and not as a result of bomb testing as hypothesized by the authors\cite{Kla04}.

Reference~\cite{Kla04} also observed lines at 2011, 2017, 2022, and 2053 keV. These lines had rates of approximately 500/(t y), 500/(t y),  300/(t y) and
380/(t y) respectively. The line at 2022 keV is near a line we predict at 2023 keV. Reference~\cite{Kla04} attributes these lines to weak transitions in
$^{214}$Bi. From our analysis it is indicated that a negligible fraction of the peak at 2023 keV is neutron-induced. However, since the predicted strength
of the tell-tale lines 
that would indicate a presence of neutron interactions is just below the sensitivity of that experiment, this conclusion is not without uncertainty.
It has been pointed out that the strength in the 2022-keV line is too strong with respect to the $^{214}$Bi branching ratios even when 
summing uncertainties are taken into account~\cite{Aal02, Bro06}. The analysis in Ref.~\cite{Aal02}, however, was based on a incorrectly normalized
Fig. 1 in Ref~\cite{HM}. A recent analysis~\cite{Bro06} taking this into account still points to an inconsistency in the line strengths.
This discrepancy could be resolved if one attributes a significant fraction of that peak
to neutron interactions on $^{76}$Ge. Such an attribution is not supported by our simulations.

Reference~\cite{cdo} simulates the background in the Heidelberg-Moscow experiment resulting in a predicted signal of 646 $\pm$ 93 counts in the
region between 2000 and 2100 keV during an exposure of 49.6 kg-y. This is a count rate of 130/(keV t y) to be compared with the quoted measured value for the 
data period simulated of 160/(keV t y). Their estimate indicates that only 0.2/(keV t y) are due to neutrons and they argue that $\mu$-generated neutrons
are a negligible contribution. Our estimates indicate that neutrons are a more significant contribution and that the $\mu$ contribution is significant. We are aware of no direct neutron flux measurements for neutrons above 25 MeV. The flux of neutrons with energy greater than 25 MeV is estimated in Ref.~\cite{cdo}  to be 10$^{-11}$/(cm$^{2}$ s) and they considered these neutrons to produce a negligible contribution to the background. In contrast, the simulation in Ref.~\cite{meihime} gives 56 $\times$ 10$^{-11}$/(cm$^{2}$ s) at 3200 m.w.e for the neutrons with energy greater than 25 MeV. We use this higher flux value and as a result, our estimate of the background rate near 2 MeV
of 50/(keV t y) is comparable to the excess (30/(keV t y)) of the measured rate in comparison to the simulated rate in Ref.~\cite{cdo}.

\subsection{Is Copper an Alternative to Lead?}
One has to consider
the existence of a DEP line at the double-beta decay endpoint a serious design consideration for Ge-detector experiments. 
From the above analysis, the dangerous lines at 2041 and 3062 keV due to Pb$(n,n'\gamma$) are not significant contributors to 
the spectrum of Ref.~\cite{Kla04}. However, as future efforts reduce the natural activity irradiating the detectors, these Pb-neutron
interactions will become important. One solution could be the use of Cu as a shield instead of Pb. Copper is rather expensive
and building the entire shield of this material is probably not necessary. A thick inner liner of Cu might suffice, but if a peak is observed
and Pb is present near the detector, arguments based on the spectrum near 3062 keV will be critical. 

Although the problematic lines we observed in the Pb data were absent in our Cu data, the shields were too dissimilar to make a quantitative
comparison regarding the effectiveness of reducing the continuum background. Furthermore, our experience with the lead and the simulation
of $(n,n'\gamma$) spectra reduces confidence in the conclusion regarding the Cu in the absence of such data. We are preparing better experimental studies to address this question.

\section{Conclusion}
\label{sec:concl}
As double-beta decay experiments become more sensitive, the potential background must be constrained to ever-lower levels.
Much progress has been made in reducing naturally-occurring radioactive isotopes from materials from which the detector
is constructed. As these isotopes that have traditionally limited the experimental sensitivity are eliminated, rarer processes
will become the dominant contributors. Here we have considered neutron-induced processes and have quantified them. Reactions
involving neutrons can result in a wide variety of contributions to the background. That is, no single component is likely to dominate.
Therefore, tell-tale signatures for neutrons are needed and were identified in this work.

In addition to the general continuum background that neutrons might produce, two specific dangerous Pb$(n,n'\gamma$) lines
were identified. These two backgrounds can be significantly reduced using depth and/or an inner layer of Cu within the shield. In particular, the 3062-keV transition in $^{207}$Pb has a double escape peak at the endpoint energy for double-beta decay in $^{76}$Ge. A comparison of past double-beta decay data indicates the rate of this transition is too small to explain a claim of double-beta decay.

\section{Acknowledgment}
We thank R.L. Brodzinski for discussions regarding the historical
production of $^{207}$Bi and its possible presence in the
environment. We thank John Wilkerson and Jason Detwiler for useful
suggestions and discussion. Finally, we
also thank Alan Poon and Werner Tornow for useful discussions and a
careful reading of the manuscript.
This work was supported in part by Laboratory Directed Research and
Development at Los Alamos National Laboratory.

%
%

\end{document}